\documentclass[12pt, draftclsnofoot, onecolumn]{IEEEtran}

\usepackage{color}
\usepackage{cite}
\usepackage{url}

\ifCLASSINFOpdf
\else
\fi
\usepackage{amsfonts,amssymb}
\usepackage{latexsym}
\usepackage{graphicx}
\usepackage{epstopdf}
\usepackage{mathrsfs}
\usepackage{amsmath}
\usepackage{caption}
\usepackage{booktabs}
\usepackage{subfigure}
\usepackage{algorithm, algorithmic}
\usepackage{bbm}
\usepackage{bm}
\usepackage{caption}
\usepackage{amsmath}
\newtheorem{theorem}{\bf{Theorem}}
\newtheorem{proposition}{\bf{Proposition}}
\newtheorem{lemma}{\bf{Lemma}}

\begin{document}
%
\title{\LARGE Deep Learning for Optimal Deployment of UAVs with Visible Light Communications}

\author{{Yining Wang,} \emph{Student Member, IEEE},  {Mingzhe Chen,} {Zhaohui Yang, Tao Luo, }\emph{Senior Member, IEEE}, and Walid Saad, \emph{Fellow, IEEE}\vspace*{-2em}\\ 
\thanks{Y. Wang and T. Luo are with the Beijing Laboratory of Advanced Information Network, Beijing University of Posts and Telecommunications, Beijing, 100876, China (e-mail: \protect\url{wyy0206@bupt.edu.cn}; \protect\url{tluo@bupt.edu.cn)}.}
\thanks{M. Chen is with the Chinese University of Hong Kong, Shenzhen, 518172, China, and also with the Department of Electrical Engineering, Princeton University, Princeton, NJ, 08544, USA (e-mail: \protect\url{mingzhec@princeton.edu}).}
\thanks{Z. Yang is Centre for Telecommunications Research, Department of Informatics, Kings College London, WC2B 4BG, UK (e-mail: \protect\url{yang.zhaohui@kcl.ac.uk}).}
\thanks{W. Saad is with the Wireless@VT, Bradley Department of Electrical and Computer Engineering, Virginia Tech, Blacksburg, VA, 24061, USA (e-mail: \protect\url{walids@vt.edu}).}
 }


\maketitle
\vspace{-0.2cm}

\begin{abstract}
In this paper, the problem of dynamical deployment of unmanned aerial vehicles (UAVs) equipped with visible light communication (VLC) capabilities for optimizing the energy efficiency of UAV-enabled networks is studied. 
In the studied model, the UAVs can simultaneously provide communications and illumination to service ground users.
Since ambient illumination increases the interference over VLC links while reducing the illumination threshold of the UAVs, it is necessary to consider the illumination distribution of the target area for UAV deployment optimization. 
This problem is formulated as an optimization problem which jointly optimizes UAV deployment, user association, and power efficiency while meeting the illumination and communication requirements of users.
To solve this problem, an algorithm that combines the machine learning framework of gated recurrent units (GRUs) with convolutional neural networks (CNNs) is proposed. 
Using GRUs and CNNs, the UAVs can model the long-term historical illumination distribution and predict the future illumination distribution. 
Given the prediction of illumination distribution, the original nonconvex optimization problem can be divided into two sub-problems and is then solved using a low-complexity, iterative algorithm.
Then, the proposed algorithm enables UAVs to determine the their deployment and user association to minimize the total transmit power.
Simulation results using real data from the Earth observations group (EOG) at NOAA/NCEI show that the proposed approach can achieve up to 68.9\% reduction in total transmit power compared to a conventional optimal UAV deployment that does not consider the illumination distribution and user association.
\end{abstract}


{\renewcommand{\thefootnote}{\fnsymbol{footnote}}
\footnotetext{A preliminary version of this work was published in the IEEE GLOBECOM 2019 \cite{wyy2019GC}.}}

%
\IEEEpeerreviewmaketitle

\section{Introduction}
Deploying unmanned aerial vehicles (UAVs) as flying base stations (BSs) for wireless networking is a flexible and cost-effective approach to providing on-demand communications \cite{MingzheChen, r195,r196, Walid1, JSAC, NanZhao}. 
However, for tomorrow's ultra dense wireless networks, UAVs deployed as aerial BSs using radio frequency (RF) will interfere with ground devices, hence significantly affecting the performance of the ground network.
In addition, the limited energy will restrict the applicability of UAVs using RF resource to provide high-speed communication services for ground users.
These challenges can be addressed by equipping UAVs with visible light communication (VLC) capabilities \cite{UAVVLC}. 
Indeed, VLC has recently attracted attention due to its large license-free bandwidth and high energy efficiency. 
For instance, a VLC system that uses light-emitting diodes (LEDs) to transmit wireless signals can provide both illumination and communication services \cite{VLCspeed}. 
Moreover, the altitude of the UAVs ensures the line of sight channel for VLC. Therefore, using VLC can be a promising approach to provide energy-efficient UAV communications with sufficiently available bandwidth. However, deploying VLC-enabled UAVs also faces many challenges that include illumination interference detection and prediction, UAV deployment optimization, and energy efficiency.

The existing literature such as in \cite{Walid1, JSAC, NanZhao} and \cite{Walid2, r193, r194, YeHu, UAV_zhaohui, r191,r192, review261} studied a number of problems related to UAV deployment. 
However, the works in \cite{Walid1, JSAC, NanZhao} ignored the energy efficiency of UAVs in optimizing the deployment of UAVs, and the works in \cite{Walid2, r193, r194} only optimized the locations of UAVs under fixed user association.
Moreover, all of the existing works such as in \cite{Walid1, JSAC, NanZhao} and \cite{Walid2, r193, r194, YeHu, UAV_zhaohui, r191,r192,review261} are over limited capacity radio frequency bands which may not allow the UAVs to meet the high data rate demands of ground users. 
Instead, VLC-enabled UAVs can be considered to provide high speed communications \cite{VLCspeed}. 
In \cite{yang}, the authors developed a novel integrated VLC and UAV framework that can simultaneously provide communication and illumination and optimized the locations of UAVs to minimize the total power consumption.
However, this work did not consider the impact of nighttime illumination such as vehicle lights, street lights, and building lights, which will cause strong interference to VLC links \cite{VLCoutdoors}. 
Therefore, it is necessary to analyze the illumination distribution of the service areas so as to optimize the deployment of VLC-enabled UAVs. 
Naturally, machine learning (ML) \cite{tutorial} can be used to predict future illumination distribution due to its strong ability on the analysis of historical illumination distribution.

More recently, there has been significant interest in applying ML techniques to optimize UAV deployment such as in \cite{ML_liu, RL_chi, review262, ANN, QianqianZhang,new_FL}.
The existing works such as in \cite{ML_liu, RL_chi, review262} used reinforcement learning (RL) algorithms to optimize network performance.
However, such works cannot be used to analyze historical illumination data and predict future illumination distribution. 
Meanwhile, the works in \cite{ANN, QianqianZhang,new_FL} only considered the temporal correlation of the network state dataset to predict future network states.
Therefore, the works in \cite{ML_liu, RL_chi, review262, ANN, QianqianZhang,new_FL} do not analyze the potential of using ML for predicting two dimensional (2D) illumination distribution which needs a comprehensive analysis of joint spatial and temporal correlations.

A number of existing works such as in \cite{conv_AE2, conv_AE3, Compositional, ConvLSTM} has studied the use of ML to capture the spatiotemporal correlations so as to predict two dimensional time-series data.
In particular, the authors in \cite{conv_AE2} and \cite{conv_AE3} studied the use of convolutional autoencoder networks (CAE) to predict time-dependent future video frame.
However, the works in \cite{conv_AE2} and \cite{conv_AE3} can only analyze the data in two consecutive time slots but ignore long-term historical information.
The authors in \cite{Compositional} presented an entity segmentation-based deep learning model to predict the image of a given scene.
However, the works in \cite{Compositional} cannot deal with the prediction of illumination distribution since it cannot be segmented by identifying the boundaries.
In \cite{ConvLSTM}, the authors used deep learning to predict the rainfall intensity in a local region.
However, the work in \cite{ConvLSTM} performed zero-padding to ensure the output is the same size as the input, which leads to blurry images.
Therefore, the solutions proposed in the prior arts \cite{conv_AE2, conv_AE3, Compositional, ConvLSTM} cannot accurately predict the illumination distribution since the boundaries and intensity of nighttime illumination caused by human activities vary in real time.
For example, during evenings, the illumination of factories will decrease while the illumination of residential or commercial areas will increase.
Meanwhile, the illumination of each road changes with the density of vehicles.
Nighttime illumination causes interference over the VLC link while reducing the illuminance requirements of users, hence affecting the data rate of each user that is serviced by VLC links and the deployment of VLC-enabled UAVs. 
Hence, it is necessary to develop a novel ML framework for the analysis and prediction of illumination distribution over ten minutes.
Based on the predictions, the network can optimally deploy UAVs to the service area for on-demand wireless service.

The main contribution of this work is a novel framework for dynamically optimizing the locations of VLC-enabled UAVs based on accurate predictions of the illumination distribution of a given area. Our key contributions include:
\begin{itemize}
\item We consider a VLC-enabled UAV network that can simultaneously provide illumination and high data rate communication services to ground users.
Compared with a conventional VLC system that uses static terminal devices, the considered VLC-enabled UAVs can avoid time-varying outdoor illumination interference.
To enhance the energy efficiency of the considered network, the UAVs must find their optimal locations and user association by predicting the distribution of ambient lighting. 
This problem is formulated as an optimization problem whose goal is to minimize the total transmit power of UAVs under illumination, communication, and user association constraints.
\item To solve this optimization problem, we propose a deep learning-based prediction approach by combining convolutional neural networks (CNNs), gated recurrent units (GRUs), and deconvolution networks (DeCNNs).
The proposed approach can analyze the temporal and spatial characteristics of the long-term historical illumination distribution.
Compared to the existing time-series prediction approach for 2D data, the proposed approach can effectively overcome the problem of blurred predictions thus enabling the UAVs to accurately predict future illumination distributions.
\item Given the predicted illumination distribution, we transform the original, nonconvex problem into a convex equivalent by using a physical relaxation for the user association constraints.
Then, we develop a feasible, efficient, and low-overhead iterative algorithm via dual decomposition, which can be implemented in VLC-enabled UAV networks.
\end{itemize}
Simulation results show that the proposed approach can achieve up to 68.9\% reduction in terms of transmit power compared to a conventional optimal UAV deployment without considering illumination distribution. 
To the best of our knowledge, this is the first work that \emph{studies the use of the predictions of the illumination distribution to provide a power-efficient deployment of VLC-enabled UAVs}.

The rest of this paper is organized as follows. The system model and the problem formulation are described in Section \ref{sec:2}. The proposed deep learning model to predict the future illumination distribution is proposed in Section \ref{sec:3}. The proposed iterative UAV deployment, user association, and power efficiency algorithm is presented in Section \ref{sec:4}. In Section \ref{sec:5}, the numerical results are discussed. Finally, conclusions are drawn in Section \ref{sec:6}.

\section{System Model and Problem Formulation}
\begin{figure}
\vspace{-0.4cm}
\centering
\setlength{\belowcaptionskip}{-0.8cm}
\includegraphics[width=9cm]{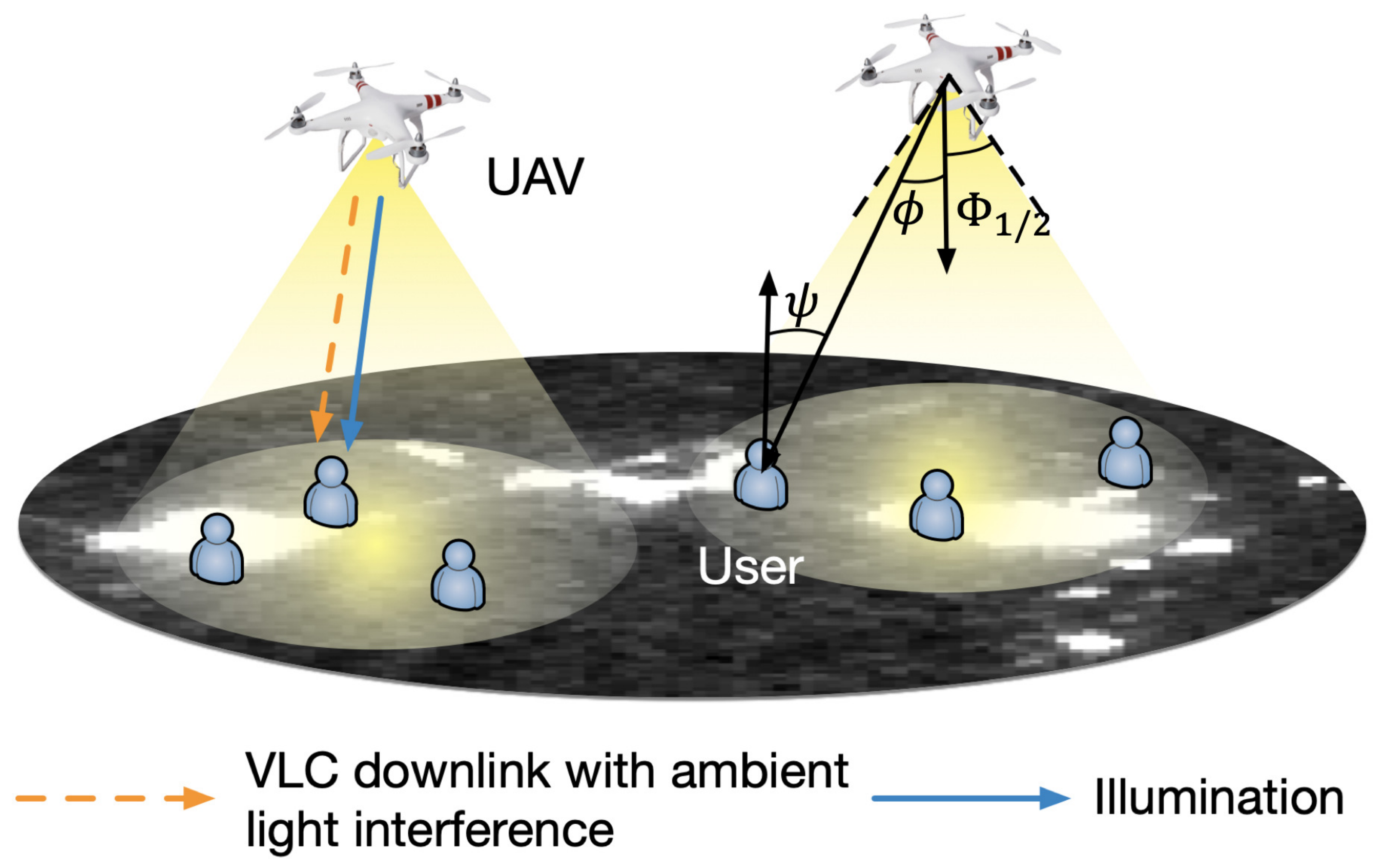}
\caption{The architecture of a cellular network that consists of UAVs and users.}
\label{fig1}
\end{figure}

Consider a wireless network composed of a set $\cal{D}$ of $\emph{D}$ VLC-enabled UAVs that serve a set $\cal{U}$ of $\emph{U}$ ground users distributed over a geographical area $\mathcal{A}$. The UAVs provide downlink transmission and illumination simultaneously, as shown in Fig. \ref{fig1}. Hereinafter, we use $\emph{aerial cell}$ to refer to the service area of each UAV.
Note that each UAV does not service ground users until it moves to the optimal location. 
Thus, during wireless transmission, the UAVs can be seen as static aerial base stations.

\label{sec:2}

\vspace{-0.1cm}
\subsection{Transmission Model}
Given a UAV $ i \in {\cal D}$ located at $\bm{q}_{i,t}=({x_i},{y_i},H)$ represents the position of UAV $i$ at time slot $t$ and a ground user $j \in {\cal U}$ located at $({v_j},{w_j}) \in {\cal A}$, the probabilistic line-of-sight (LoS) and non-line-of-sight (NLoS) channel model is used to model the VLC link between UAV $i$ and ground user $j$ \cite{JSAC}.
For simplicity, we do not consider the diffusion of visible light in outdoor environments. 
Therefore, the LoS and NLoS channel gain of the VLC link between UAV $i$ and user $j$ can be given by \cite{KomineFundamental}:
\vspace{-0.1cm}
\begin{equation}\label{los}
\begin{aligned}
{h_j^{\rm{LoS}}} &= {\!\frac{{(m + 1)\rho}}{{2\pi d_{ij}^2}}g(\psi ){{\cos }^m}(\phi )\cos \left( \psi  \right),\;0\leqslant \psi \leqslant \Psi_c,}\\
{h_j^{\rm{NLoS}}} &= 0
\end{aligned}
\end{equation}
\vspace{-0.2cm}
\begin{equation}\label{nlos}
\vspace{-0.1cm}
{h_j^{\rm{NLoS}}} = 0,
\end{equation}
where $\rho$ is the detector area and $d_{ij} = \sqrt {{{({v_j} - {x_i})}^2} + {{({w_j} - {y_i})}^2} + H^2} $ is the distance between UAV $i$ and ground user $j$. $m =  - \ln 2/\ln (\cos {\Phi _{1/2}})$ is the Lambert order with ${\Phi _{1/2}}$ being the transmitter semiangle (at half power); $\psi$ and $\phi$ represent the angle of incidence and irradiance, respectively. As such, $\cos \phi  = \cos \psi  = \frac{H}{d_{ij}}$. Let ${{\Psi _c}}$ be the receiver field of vision (FOV) semi-angle.
The gain of the optical concentrator $g(\psi )$ is defined as:
\begin{equation}
g(\psi ) = \left\{ {\begin{array}{*{20}{l}}
{\frac{{{n_e}^2}}{{{{\sin }^2}{\Psi _c}}},\;0 \leqslant {\psi} \leqslant {\Psi _c},}\\
{\;\;\;\;0\;\;\;,\;{\psi} > {\Psi _c}},
\end{array}} \right.
\end{equation}
where $n_e$ represents a refractive index.
According to \cite{JSAC}, the probability of the LoS link will be: $B \left({h_j^{\rm{LoS}}}\right) = {(1 + X\exp ( - Y[{\tau _j} - X]))^{ - 1}}$, where $X$ and $Y$ are environmental parameters and ${\tau _j}=\sin^{-1}(H/d_{ij})$ is the elevation angle.
Clearly, the average channel gain from UAV $i$ to user $j$ can be given by:
\vspace{-0.1cm}
\begin{equation}\label{average}
\vspace{-0.1cm}
{h_j(x_i,y_i)}=B \left({h_j^{\rm{LoS}}}\right) \times {h_j^{\rm{LoS}}}+B \left({h_j^{\rm{NLoS}}} \right) \times {h_j^{\rm{NLoS}}},
\end{equation}
where $B \left({h_j^{\rm{NLoS}}} \right) = 1- B \left({h_j^{\rm{LoS}}}\right)$.

Let $u_{ij,t}$ be the association for UAV $i$ and user $j$ at time $t$, i.e., $u_{ij,t}=1$ indicates that user $j$ is associated with UAV $i$ at time $t$; otherwise, we have $u_{ij,t}=0$. Assuming that each user is associated with only one UAV, we have:
\vspace{-0.1cm}
\begin{equation}
\sum\limits_{i \in \cal{D}} {{u_{ij,t}}}  = 1,\forall j \in \cal{U}.
\end{equation}
For static user $j$ located at $\left(v_j,w_j\right)$ associated with UAV $i$, the channel capacity at time $t$ can be given by:
\begin{equation}
{C_{ij,t}} = \frac{1}{2}{\log _2}\left( {1 + \frac{e}{{2\pi }}{{\left( {\frac{{\xi {P_{ij,t}}{h_j}(x_i,y_i)}}{{{n_w} + {I_t}(v_j,w_j)}}} \right)}^2}} \right),
\label{eq:capacity}
\end{equation}
where $\xi$ is the illumination target, $P_{ij,t}$ is the transmit power of UAV $i$ serving user $j$ at time $t$, and ${n_w}$ represents the standard deviation of the additive white Gaussian noise. 
In (\ref{eq:capacity}), ${I_t}(v_j,w_j)$ is the ambient illumination at $(v_j,w_j)$, which also indicates the interference over the VLC link between the UAV and the user $j$.
To obtain the illumination for each location, we define the illumination distribution of the service area as ${\bm{I}}_t$ that will be specified in Section \ref{sec:3}. 

Due to the limited energy of UAVs, their deployment must be optimized to minimize the transmit power while satisfying the data rate and illumination requirements of users. 
Since the area of aerial cells is small and ground users served by UAVs are static, as done in \cite{mobile2}, we do not consider the mobility energy consumption of the UAVs. 

\vspace{-0.2cm}
\subsection{Problem Formulation}
To formulate the deployment problem, we must first determine the minimum transmit power that each UAV $i$ uses to meet the data rate and illumination requirements of its associated users. 
Then, substituting (\ref{average}) into (\ref{eq:capacity}), the power required for UAV $i$ to satisfy the data rate requirement $R_j$ of user $j$ will be:
\begin{equation} \label{power_p}
P_{ij,t} = \frac{{u_{ij,t}({n _w} + {I_t}(v_j,w_j))\sqrt {\frac{{2\pi }}{e}({2^{2{R_j}}} - 1)} }}{{\xi B \left({h_j^{\rm{LoS}}}\right) {h_j^{\rm{LoS}}}}},
\end{equation}
From (\ref{power_p}) we can see that, the power required to satisfy the data rate requirement of user $j$ depends on both the position of associated UAV $i$ and the LoS probability $B \left({h_j^{\rm{LoS}}}\right)$.
This makes (\ref{power_p}) difficult to handle for the purpose of UAV deployment optimization.
As done in \cite{rui_los7}, we use a homogeneous approximation for the LoS probability, i.e. by letting $B \left({h_j^{\rm{LoS}}}\right) \approx \bar{B}, \forall j \in {\cal U}$.
$\bar{B}$ is set as the average value based on a certain UAV deployment.
Given $\bar{B}$, the power required for UAV $i$ to satisfy the data rate constraint of user $j$ is:
\begin{equation} \label{power}
P_{ij,t} = \frac{{u_{ij,t}({n _w} + {I_t}(v_j,w_j))\sqrt {\frac{{2\pi }}{e}({2^{2{R_j}}} - 1)} }}{{\xi \bar{B} {h_j^{\rm{LoS}}}}}.
\end{equation}
A UAV can successfully satisfy all the users’ requirements once the user who has the maximum power requirement is satisfied. Therefore, the minimum transmit power of UAV $i$ satisfying the data rate requirements of its associated users is given by:
\vspace{-0.1cm}
\begin{equation}\label{power_min}
P_{i,t}^\textrm{min} = \max \{{P_{ij,t}}\} ,\forall j \in {{\cal U}}.
\end{equation}

Given this system model, our goal is to find an effective deployment of UAVs that meets the data rate and illumination requirements of each user while minimizing the transmit power of the UAVs. This problem involves predicting the illumination and adjusting the user association, the locations as well as the transmit powers of UAVs, which is given by:
\begin{subequations}\label{optimal_problem}
\begin{align}\tag{\theequation}
&{\mathop {\min }\limits_{{x_{i,t}}, {y_{i,t}}, \bm{u}_{i,t}} \;\;\sum\limits_{i \in {\cal{D}}} {\left( {P_{i,t}} \right)} ,}\\
&{\;\;\;\;\;{\rm{s}}.{\rm{t}}.\;\;\;\;\;{\xi {P_{i,t}}{h_j}(x_{i,t},y_{i,t})} \geqslant  u_{ij,t}(\eta _r - {I_t(v_j,w_j)} ), \;\forall i \in {\cal{D}}, \forall j \in \cal{U}, }\\
&{\;\;\;\;\;\;\;\;\;\;\;\;\;\;\;{{P}_{i,t}} \geqslant P_{i,t}^{\min }, \;\forall i  \in  {\cal{D}},}\\
&{\;\;\;\;\;\;\;\;\;\;\;\;\;\;\;\sum\limits_{i \in \cal{D}} {{u_{ij,t}}}  = 1,\;\forall j \in {\cal{U}}, }\\
&{\;\;\;\;\;\;\;\;\;\;\;\;\;\;\;{u_{ij,t}} \in \{ 0,1\},\;\forall i \in {\cal{D}}, \forall j \in \cal{U}, }\\
&{\;\;\;\;\;\;\;\;\;\;\;\;\;\;\;{\left\| {\bm{q}_{i,t} - \bm{q}_{k,t}} \right\|^2} \geqslant {d_{\min }},\;\forall i,k \in {\cal{D}}, i \ne k,}
\end{align}
\end{subequations}
where $P_{i,t}$ is the transmit power of UAV $i$ at time $t$, $\bm{u}_{i,t}=\left[u_{i1,t},u_{i2,t},\ldots,u_{iU,t}\right]$ is the user association vector of UAV $i$, ${\eta _r}$ denotes the illumination demand, ${\xi {P_{i,t}}{h_i}(x_i,y_i)}$ is the illumination of UAV $i$ at time $t$, and ${d_{\min }}$ is the minimum distance between any two UAVs.
(\ref{optimal_problem}a) indicates that each UAV $i$ needs to provide illumination to meet the illumination threshold of its associated users. 
(\ref{optimal_problem}b) indicates that the transmit power of UAV $i$ should satisfy the data rate requirements of its associated users from (\ref{power_min}). (\ref{optimal_problem}c) and (\ref{optimal_problem}d) imply that each user can only associate with one UAV at each time slot. 
(\ref{optimal_problem}e) guarantee the service area of each UAV does not overlap with the service areas of other UAVs, and, hence, we ignore the interference caused by other UAVs.
Note that ambient illumination causes interference over the VLC link while reducing the illuminance requirements of users.
The distribution of ambient illumination at night that consists of vehicle, street, and building lights varies in real time.
For example, during nights, the illumination of factories will decrease while the illumination of residential or commercial areas will increase.
In addition, the illumination of each road changes as the vehicle density changes.
Therefore, it is necessary to predict the illumination distribution of the target area to deploy the UAVs at the beginning of each time interval. 
We next introduce a machine learning algorithm to predict the illumination distribution of the service area over ten minutes and, then, deploy the UAVs based on the solution of (\ref{optimal_problem}). 

\section{Machine Learning for Illumination Prediction}
\label{sec:3}
\begin{figure}
\centering
\setlength{\belowcaptionskip}{-0.5cm}
\vspace{-0.4cm}
\includegraphics[width=11.5cm]{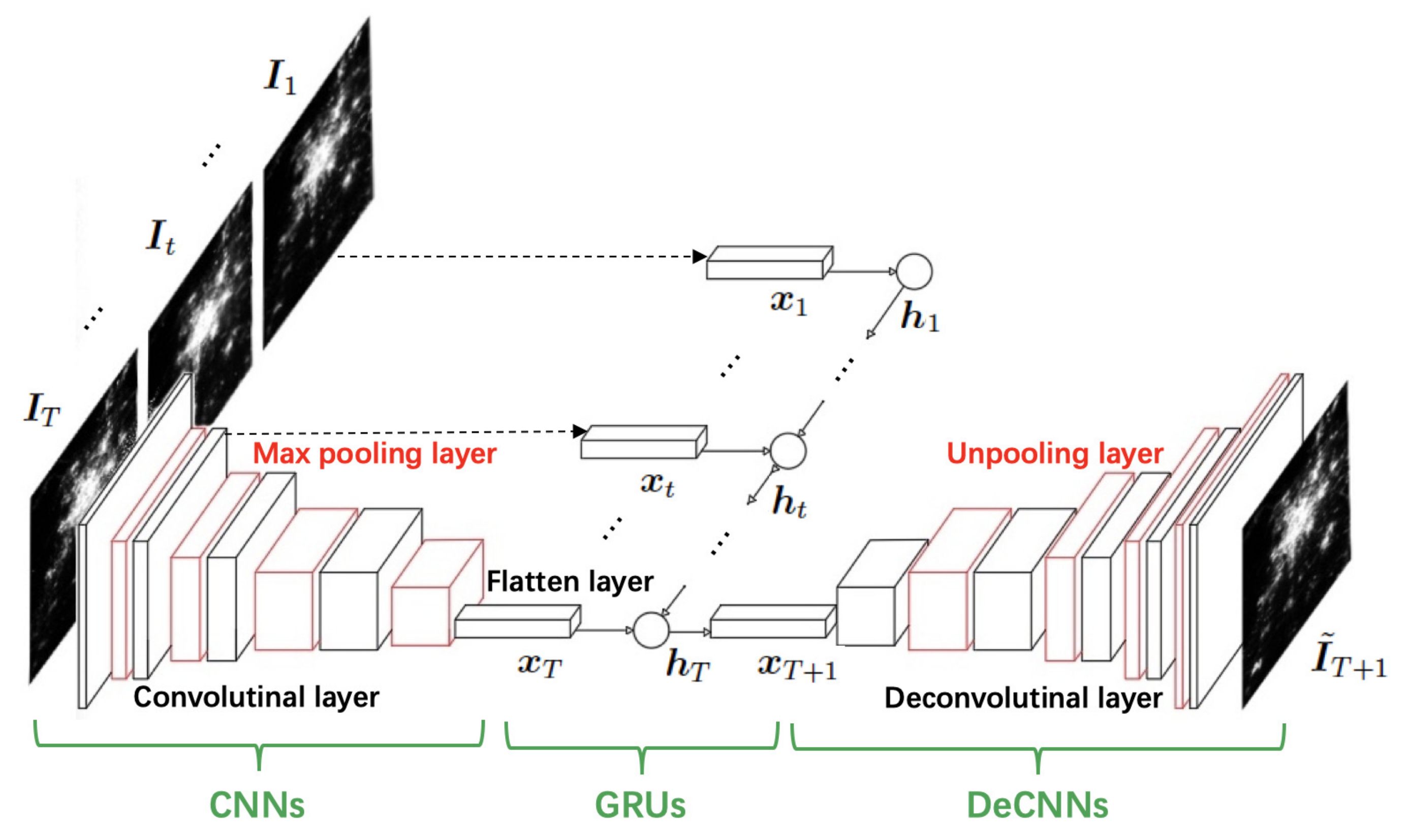}
\caption{Overall architecture of the proposed learning model.}
\label{fig6}
\vspace{-0.3cm}
\end{figure}

Since predicting the illumination distribution requires both spatial and temporal sequence information, we propose a deep learning approach that integrates GRUs with CNNs. 
The proposed approach enables the UAVs to analyze the relationship among historical illumination distributions for the future illumination distribution prediction.
We first apply an CNN to extract spatial features of the illumination distribution at each time slot $t$.
Then, the time-varying spatial features are fed to GRUs for predicting the features of illumination distribution at time $t+1$ based on the learned temporal dependencies. 
Finally, a deconvolution network (DeCNN) is used to transform the multidimensional features, which are predicted by GRUs, to the illumination distribution.
The architecture of the integrated GRU and CNN predictive model is shown in Fig.~\ref{fig6}.
Next, we introduce the three components of our model: a) CNNs, b) GRUs, and c) DeCNNs.
\vspace{-0.3cm}
\subsection{CNN for Encoding Illumination Distribution}
Since illumination is caused by human activities such as business and industrial operation, the illumination at a given position always has very strong spatial correlations with the illumination distribution of nearby regions. 
Therefore, we use CNNs to capture spatial correlations between the illumination of a given location and the illumination of its nearby regions, and then build the feature representations that preserve the changes in local illumination.

Given an illumination distribution $\boldsymbol{I}_t$ at time $t$, a CNN encoder is used to extract the feature vector $\bm{x}_t$, which represents the spatial features extracted from $\boldsymbol{I}_t$.
The proposed CNN algorithm consists of $L$ convolutional layers, $L$ max-pooling layers, and a flatten layer.
In particular, each convolutional layer is followed by a max-pooling layer and the last layer of the CNN is a flatten layer. 
Next, we introduce each layer of the proposed CNN.
\begin{itemize}
\item \emph{Convolutional layer}: 
In a CNN, a convolutional layer is used to extract spatial features which are useful in the next illumination distribution predicting stage.
Mathematically, the input of each convolutional layer $l$ is $\boldsymbol{H}_t^{l-1,m}$, where $\boldsymbol{H}_t^{l-1,m} , l=1, \cdots, L$ is the feature map $m$ in convolutional layer $l-1$ and the input ${\boldsymbol{H}_t^{0,1}}$ of convolutional layer $1$ is an illumination distribution at time $t$ (i.e., $\boldsymbol{I}_t = {\boldsymbol{H}_t^{0,1}}$). 
The output of each convolutional layer $l$ is given by:
\vspace{-0.1cm}
\begin{equation}\label{convolution}
\boldsymbol{H}_t^{l,m} = f(\sum\limits_{k =1}^{K_c^{l-1}} {{\boldsymbol{H}}_t^{l - 1,k} \otimes {\boldsymbol{W}_{c,t}^{l,m}} }+{\bm{b}_{c,t}^{l,m}}),
\end{equation}
where $f(\cdot)= \max(0,\cdot)$ is rectifier activation function, $K_c^{l-1}$ is the number of feature maps in convolutional layer $l-1$, $\otimes$ denotes the convolution operation, and ${\boldsymbol{W}_{c,t}^{l,m}} \in {\mathbb{R}^{S \times S}}$ and ${\bm{b}_{c,t}^{l,m}}$ are convolution kernels and bias of feature map $m$ in convolutional layer $l$, respectively, with $S$ being a constant that controls the spatial granularity. 
Note that, for each feature map $\boldsymbol{H}_t^{l,m} \in {\mathbb{R}^{\lambda_l \times \lambda_l}}$, the size $\lambda_{l}$ of the feature map satisfies $\lambda_{l}= \lambda_{l-1}-S+1$ and ${\boldsymbol{H}_t^{0,0}} \in {\mathbb{R}^{\lambda_0 \times \lambda_0}}$.

\item \emph{Max-pooling layer}:
The input of each max-pooling layer $l$ is the feature map $\boldsymbol{H}_t^{l,m}$. 
Max-pooling layers compress the input feature maps, which allow a CNN encoder to extract robust spatial features while reducing the computation complexity.
The positions of max-pooled features in feature maps are recorded in switch variables (switches), which will be used to decode the predicted features of the future illumination distribution in the DeCNN.

\item \emph{Flatten layer}:
A flatten layer is used at the end of the CNN encoder, whose input is the combination of feature maps extracted by max-pooling layer $L$.
The flatten layer generate a spatial feature vector $\bm{x}_t \in {\mathbb{R}^N}$, where $N={\lambda_L}^2{K_L}$ is the number of the features extracted by the CNN encoder.
\end{itemize} 

\vspace{-0.3cm}
\subsection{Illumination Distribution Prediction}
Next, we introduce the use of GRUs \cite{GRU} for the prediction of the illumination distribution. GRUs are extensions of conventional recurrent neural networks (RNNs)\cite{tutorial}. GRUs can effectively solve the gradient vanishing and the gradient exploding problem in long-term memory RNNs. Due to interconnected neurons at hidden layers and their internal gating mechanisms, GRUs can model the temporal characteristics of the long-term illumination distribution. In addition, GRUs can dynamically update the model based on the current illumination distribution due to the variable-length recurrent structure, hence, GRUs enable the UAVs to predict future illumination distribution.

A GRU-based prediction algorithm consists of three components: a) input, b) output, and c) GRU model. The key components of our GRU-based prediction approach are:
\begin{itemize}
\item \emph{Input}: The input of the GRU is the output of the CNN encoder which is represented as $\boldsymbol{X} = \left(\bm{x}_1, \bm{x}_2,\cdots ,\bm{x}_t, \cdots, \bm{x}_T\right)$.
\item \emph{Output}: The output of the GRU-based prediction algorithm is a vector $\bm{x}_{T+1}$, that represents the spatial features of illumination distribution at time slot $T+1$.
\item \emph{GRU model}:  A GRU model is used to approximate the function between the input $\boldsymbol{X}$ and output $\bm{x}_{T+1}$, thus building a relationship between historical illumination distribution and future illumination distribution. A GRU model is essentially a dynamic neural network that consists of an input layer, a hidden layer, and an output layer. The hidden states $\bm{h}_t$ of the units of the in hidden layer at time $t$ are used to store information related to the illumination distribution from time slot $1$ to $t$. For each time $t$, the hidden states $\bm{h}_t$ of the GRU are updated based on the input $\bm{x}_t$ and $\bm{h}_{t-1}$. Next, we introduce how to update the hidden state $h_t^j$ of hidden unit $j$ given a new illumination distribution $\bm{x}_t$.

\setlength\parindent{1em}At each time slot $t$, the hidden state $h_t^j$ is determined by two gates: reset gate $r_t^j$ and update gate $z_t^j$. First, the reset gate $r_t^j$ is used to determine the historical illumination distribution information retained in the candidate hidden state ${\tilde{h}_t^j}$, which can be given by:
\begin{equation}\label{reset_gate}
\setlength{\abovedisplayskip}{3pt}
\setlength{\belowdisplayskip}{3pt}
r_t^j = \sigma ({[{{\boldsymbol{W}}_r}{\bm{x}_t}]_j} + {[{{\boldsymbol{U}}_r}{\bm{h}_{t - 1}}]_j}),
\end{equation}
where $\sigma (\cdot)=\frac{1}{{1 + {e^{ - (\cdot)}}}}$ is the logistic sigmoid function and ${[ \cdot ]_j}$ is element $j$ of a vector. ${{\boldsymbol{W}}_r} \in {\mathbb{R}^{{N} \times {D_h}}}$ and ${{\boldsymbol{U}}_r} \in {\mathbb{R}^{{D_h} \times {D_h}}}$ represent the weight matrices of reset gate, where $N$ is the length of the input ${\bm{x}_t}$ and $D_h$ is the number of the units in hidden layer. Based on the value of the reset gate $r_t^j$, the candidate hidden state ${\tilde{h}_t^j}$ that is used to combine the input illumination distribution $\bm{x}_t$ with the previous memory ${\bm{h}_{t - 1}}$ is given by:
\begin{equation}\label{reset}
\setlength{\abovedisplayskip}{3pt}
\setlength{\belowdisplayskip}{3pt}
{\tilde{h}_t^j} = \tanh \left({[{{\boldsymbol{W}}_{\tilde{h}}}{\bm{x}_t}]_j} + {[{{\boldsymbol{U}}_{\tilde{h}}}({\bm{r}_t} \odot {\bm{h}_{t - 1}})]_j}\right),
\end{equation}
where ${\bm{r}_t} \in  {\mathbb{R}^{D_h}}$ is a reset gate vector at time $t$ and $\odot$ is an element-wise multiplication. For example, given two vectors ${\bm{p}}=(a,b)$ and ${\bm{q}}=(c,d)$, ${\bm{p}} \odot {\bm{q}}=(ac,bd)$. ${{\boldsymbol{W}}_{\tilde{h}}} \in {\mathbb{R}^{{N} \times {D_h}}}$ and ${{\boldsymbol{U}}_{\tilde{h}}} \in {\mathbb{R}^{{D_h} \times {D_h}}}$ represent the hidden state weight matrices.

Similarly, the update gate ${z_t^j}$ is used to decide the size of the information stored in the candidate hidden state to update the hidden state $h_t^j$, which can be given by:
\begin{equation}\label{update_gate}
\setlength{\abovedisplayskip}{3pt}
\setlength{\belowdisplayskip}{3pt}
{z_t^j} = \sigma \left({[{{\boldsymbol{W}}_z}{\bm{x}_t}]_j} + {[{{\boldsymbol{U}}_z}{\bm{h}_{t - 1}}]_j}\right),
\end{equation}
where ${{\boldsymbol{W}}_z} \in {\mathbb{R}^{{N} \times {D_h}}}$ and ${{\boldsymbol{U}}_z} \in {\mathbb{R}^{{D_h} \times {D_h}}}$ represent the weight matrices of the update gate. The actual hidden state $h_t^j$ of hidden unit $j$ is updated by:
\begin{equation}\label{hidden_unit}
\setlength{\abovedisplayskip}{3pt}
\setlength{\belowdisplayskip}{3pt}
h_t^j =  {z_t^j}{h_{t-1}^j} + (1 - {z_t^j}){\tilde{h}_t^j}.
\end{equation}

The proposed GRU model iteratively updates the hidden states to store the input $\boldsymbol{X}$ until the hidden state of the current time $T$ is computed. The output layer of the GRU model will predict the illumination distribution at time ${T+1}$ based on the hidden state ${\bm{h}_T}$:
\begin{equation}\label{output}
\setlength{\abovedisplayskip}{3pt}
\setlength{\belowdisplayskip}{3pt}
{\bm{x}}_{T+1} =  {{\boldsymbol{W}}_o}{\bm{h}_T},
\end{equation}
where ${\boldsymbol{W}}_o \in {\mathbb{R}^{{D_h} \times {N}}}$ is the output weight matrix. 
Based on (\ref{output}), we get output $\bm{x}_{T+1}$ from the hidden state ${\bm{h}_T}$ that stores the information of input $\boldsymbol{X}$.
\end{itemize}

\vspace{-0.3cm}
\subsection{Illumination Distribution Deconvolution Network}
We now study the decoding of the predicted feature vector ${\bm{x}_{T+1}}$ into the illumination distribution $\boldsymbol{I}_{T+1}$.
Since GRU-based predictions ${\bm{x}_{T+1}}$ only contain the spatial features of illumination distribution $\boldsymbol{I}_{T+1}$ rather than a complete illumination distribution, we use a DeCNN to decode the predicted features.
The proposed DeCNN decoder is a mirrored version of the CNN encoder introduced before, which consists of $L$ unpooling layers and $L$ deconvolutional layers. Next we introduce each layer of the proposed DeCNN.
\begin{itemize}
\item \emph{Unpooling layer}:
The input of the first unpooling layer is ${\bm{x}}_{T+1}$ predicted by GRUs and the input of unpooling layer $l$ ($l>1$) is the feature maps output from the deconvolutional layer $l-1$.
The unpooling layers are used to reconstruct the illumination distribution of service area to the original size.
Therefore, the output of an unpooling layer is an enlarged, yet sparse feature map.

\item \emph{Deconvolutional layer}: 
The input of each deconvolutional layer $l$ is the enlarged feature maps output from the unpooling layer $l-1$.
The deconvolutional layers effectively reconstruct the detailed structure of illumination distribution based on the learned weights, which is given by:
\begin{equation}\label{deconvolution}
\vspace{-0.2cm}
\tilde{\boldsymbol{H}}_{T+1}^{l,m} = f(\sum\limits_{k=1}^{K_d^{l-1}} {\tilde{\boldsymbol{H}}_{T+1}^{l - 1,k} \otimes {\boldsymbol{W}_{d,T+1}^{l,m}} }+{\bm{b}_{d,T+1}^{l,m}}),
\end{equation}
where $\tilde{\boldsymbol{H}}_{T+1}^{l,m}$ is reconstructed feature map $m$ in deconvolutional layer $l$, $K_d^{l-1}$ is the number of feature maps in deconvolutional layer $l-1$, and ${\boldsymbol{W}_{d,T+1}^{l,m}}$ and ${\bm{b}_{d,T+1}^{l,m}}$ are convolution kernels and bias of feature map $m$ in deconvolutional layer $l$, respectively. 
The output of the last deconvolutional layer is $\tilde{\boldsymbol{H}}_{T+1}^{L} \in {\mathbb{R}^{{\lambda_0} \times {\lambda_0}}}$, equivalent to $\tilde{\boldsymbol{I}}_{T+1}$, which represents the prediction of illumination distribution at time $T+1$.
\end{itemize}

Finally, the trained integrated GRU and CNN predictive model can output the illumination distribution prediction $\tilde{\boldsymbol{I}}_{T+1}$ based on the input historical illumination distributions.
Once $\tilde{\boldsymbol{I}}_{T+1}$ is obtained, given a user located at $(v_j,w_j)$ where $v_j \in \{0, 1, \cdots, {\lambda_0}\}$ and $w_j \in \{0, 1, \cdots, {\lambda_0}\}$, the ambient illumination $I_{T+1}(v_j,w_j)$ can be obtained. 
Then, the ambient illumination $I_{T+1}(v_j,w_j)$ of each user $j$ can be substituted into the optimization problem (\ref{optimal_problem}).

\begin{algorithm}[t]
\scriptsize
\caption{Integrated GRU and CNN Predictive Model for Illumination Distribution Prediction.}
\begin{algorithmic}[1]
\STATE \textbf{Input:} The time series illumination distribution of service area, $\boldsymbol{I}_1, \boldsymbol{I}_2, \cdots, \boldsymbol{I}_t, \cdots, \boldsymbol{I}_T$.
\STATE \textbf{Initialize:} ${\boldsymbol{W}_{c,1}}, \cdots, {\boldsymbol{W}_{c,T}}, {\boldsymbol{W}_{d,T+1}}, {\boldsymbol{W}_r}, {\boldsymbol{U}_r}, {\boldsymbol{W}_z}, {\boldsymbol{U}_z}, {{\boldsymbol{W}}_{\tilde{h}}}, {{\boldsymbol{U}}_{\tilde{h}}}$, and ${\boldsymbol{W}_o}$ are initially generated randomly via a uniform distribution. The number of iterations $e$.
\FOR {$i = 1 \to e$}
\FOR {each time $t$}
\STATE Input $\boldsymbol{I}_t$ and encode $\boldsymbol{I}_t$ into a feature vector ${\bm{x}}_{t}$ based on (\ref{convolution}).
\STATE Predict the spatial feature vector ${\bm{x}}_{t+1}$ based on (\ref{output}).
\STATE Decode the predicted ${\bm{x}}_{t+1}$ into the illumination distribution $\boldsymbol{I}_{t+1}$.
\ENDFOR
\STATE Calculate the loss $E$ based on (\ref{loss}).
\STATE Update the weight matrices based on (\ref{weight}).
\ENDFOR
\STATE \textbf{Output:} Prediction ${\boldsymbol{I}_{T+1}}$.
\end{algorithmic}
\label{algorithm_1}
\end{algorithm}

\vspace{-0.3cm}
\subsection{Integrated GRU and CNN Predictive Model Training}
The proposed integrated GRU and CNN predictive model build the relationship between output $\tilde{\boldsymbol{I}}_{T+1}$ and the input time series historical illumination distribution $\boldsymbol{I}_1, \boldsymbol{I}_2, \cdots, \boldsymbol{I}_t, \cdots, \boldsymbol{I}_T$ using the weight parameters.
To build this relationship, a batch gradient descent approach is used to train the weight matrices which are initially generated randomly via a uniform distribution.
The update rule of the gradient descent approach is given by:
\begin{equation}\label{weight}
\setlength{\abovedisplayskip}{3pt}
\setlength{\belowdisplayskip}{3pt}
{\boldsymbol{W}}_n^{i+1} = {\boldsymbol{W}}_n^i - \alpha \nabla E({\boldsymbol{W}}_n), \;\;{\boldsymbol{U}}_m^{i+1} = {\boldsymbol{U}}_m^i - \alpha \nabla E({\boldsymbol{U}}_m),
\end{equation}
where $\alpha$ is the learning rate, $n \in \left \{ c, d, r, z, \tilde{h}, o \right \}$, and $m \in \left\{r, z, \tilde{h} \right\}$. $\nabla E({\bf{W}}_n) = \frac{{\partial E}}{{\partial {{\boldsymbol{W}}_n}}}$ and $\nabla E({\boldsymbol{U}}_m) = \frac{{\partial E}}{{\partial {{\boldsymbol{U}}_m}}}$ are the gradients of the loss function $E$ which is defined as:
\begin{equation}\label{loss}
\setlength{\abovedisplayskip}{4pt}
\setlength{\belowdisplayskip}{4pt}
E = \frac{1}{{2{\lambda_0}^2}}\sum\limits_{x = 1}^{{\lambda _0}} {\sum\limits_{y = 1}^{{\lambda _0}} {{||{\boldsymbol{I}_{T+ 1}(x,y)} - {{\tilde {\boldsymbol{I}}}_{T + 1}(x,y)}||}^2} }.
\end{equation}
${\boldsymbol{I}_{t + 1}^n} $ and ${{\tilde {\boldsymbol{I}}}_{t + 1}^n}$ represent the actual illumination and the predicted illumination at location $n$ at time $t+1$, respectively. 
The specific process of using the proposed deep learning model to predict the illumination distribution for each UAV $i$ is summarized in \textbf{Algorithm 1}.

\vspace{-0.2cm}
\section{Optimization of UAV Deployment, User Association, and Power Efficiency}
\label{sec:4}
Once the illumination distribution is predicted, the UAVs can determine their optimal deployment at the beginning of each time interval by solving the optimization problem defined in (\ref{optimal_problem}). 
We assume that the positions of UAVs remain unchanged during each prediction period.
As analyzed in Section \ref{sec:2}, a UAV only needs to consider the user with the maximum power requirement since, by doing so, the requirements of all other users will be automatically satisfied. Therefore, substituting (\ref{average}), (\ref{power}), and (\ref{power_min}) into (\ref{optimal_problem}), we have:
\begin{subequations}\label{optimal_problem1}
\begin{align}\tag{\theequation}
&{\mathop {\min \;}\limits_{{x_{i,T+1}},{y_{i,T+1}},P_{i,T + 1},{\bm{u}}_{T+1}} \;\sum\limits_{i \in {\cal{D}}} {{P_{i,T + 1}}} ,}\\
&{\;\;\;\;\;\;\;\;\;\;\;\;\;\;\;{\rm{s}}.{\rm{t}}.\;\;\;\;\;\;\;\;\;\;\;\;\;{P_{i,T + 1}} \geqslant lM_j{d_{ij}^{m+3}}u_{ij,T+1},\; \forall i \in {\cal{D}},\forall j \in {{\cal{U}}},}\\
&{\;\;\;\;\;\;\;\;\;\;\;\;\;\;\;\;\;\;\;\;\;\;\;\;\;\;\;\;\;\;\;\;\;{P_{i,T + 1}} \geqslant lN_j{d_{ij}^{m+3}}u_{ij,T+1}, \;\forall i \in {\cal{D}},\forall j \in {{\cal{U}}},}\\
&{\;\;\;\;\;\;\;\;\;\;\;\;\;\;\;\;\;\;\;\;\;\;\;\;\;\;\;\;\;\;\;\;\;\sum\limits_{i \in D} {{u_{ij,T + 1}} = 1}, \;\forall j \in {\cal{U}}, }\\
&{\;\;\;\;\;\;\;\;\;\;\;\;\;\;\;\;\;\;\;\;\;\;\;\;\;\;\;\;\;\;\;\;\;{u_{ij,T + 1}} \in \{ 0,1\},\;\forall i \in {\cal{D}},\forall j \in {{\cal{U}}}, }\\
&{\;\;\;\;\;\;\;\;\;\;\;\;\;\;\;\;\;\;\;\;\;\;\;\;\;\;\;\;\;\;\;\;\;{\left\| {\bm{q}_{i,T+1} - \bm{q}_{k,T+1}} \right\|^2} \geqslant {d_{\min}},\;\forall i,k \in {\cal{D}}, i \ne k,}
\end{align}
\end{subequations}
where $l=\frac{2\pi}{{\xi (m + 1) \rho  g(\psi ) H^{m+1} }}$, $M_j={{\eta _r} - {I_{T + 1}(v_j,w_j)}}$, and $N_j={({n_w} + {I_{T + 1}(v_j,w_j)})\sqrt {\frac{{2\pi }}{e}({2^{2{R_j}}} - 1)}}$.
Note that problem (\ref{optimal_problem1}) is nonconvex. We present an iterative algorithm for solving the nonconvex problem.
In particular, we first optimize the UAV deployment and power allocation with fixed user association. 
Then, given the UAV deployment, we find the optimal user association. 

\vspace{-0.3cm}
\subsection{UAV Deployment and Power Efficiency with Fixed User Association}
Since constraints (\ref{optimal_problem1}c) and (\ref{optimal_problem1}d) are only determined by user association ${\bm{u}}_{T+1}$, the UAV deployment and power efficiency problem (\ref{optimal_problem1}) with fixed user association ${\bm{u}}_{T+1}$ is expressed as: 
\begin{subequations}\label{optimal_problem2}
\setlength{\abovedisplayskip}{3pt}
\setlength{\belowdisplayskip}{3pt}
\begin{align}\tag{\theequation}
&{\mathop {\min \;}\limits_{{x_{i,T+1}},{y_{i,T+1}},P_{i,T + 1}} \;\sum\limits_{i \in {\cal{D}}} {{P_{i,T + 1}}} ,}\\
&{\;\;\;\;\;\;\;\;\;\;{\rm{s}}.{\rm{t}}.\;\;\;\;\;\;\;\;\;\;{P_{i,T + 1}} \geqslant lM_j{d_{ij}^{m+3}},\; \forall i \in {\cal{D}},\forall j \in {{\cal{U}}_i},}\\
&{\;\;\;\;\;\;\;\;\;\;\;\;\;\;\;\;\;\;\;\;\;\;\;\;\;{P_{i,T + 1}} \geqslant lN_j{d_{ij}^{m+3}}, \;\forall i \in {\cal{D}},\forall j \in {{\cal{U}}_i},}\\
&{\;\;\;\;\;\;\;\;\;\;\;\;\;\;\;\;\;\;\;\;\;\;\;\;\;{\left\| {\bm{q}_{i,T+1} - \bm{q}_{k,T+1}} \right\|^2} \geqslant {d_{\min}},\;\forall i,k \in {\cal{D}}, i \ne k,}
\end{align}
\end{subequations}
where ${{\cal{U}}_i}=\{j \in{\cal{U}} |u_{ij,T+1}=1\}$.
In constraint (\ref{optimal_problem2}c), since ${\left\| {\bm{q}_{i,T+1} - \bm{q}_{k,T+1}} \right\|^2}$ is a convex function with respect to $\bm{q}_{i,T+1}$ and $\bm{q}_{k,T+1}$, we have the following inequality by applying the first-order Taylor expansion at the given point $\bm{q}^{(r)}_{i,T+1}$ and $\bm{q}^{(r)}_{k,T+1}$:
\begin{equation}\label{Taylor}
{\left\| {\bm{q}_{i,T+1} \!-\! \bm{q}_{k,T+1}} \right\|^2} \!\geqslant\! \!-\!{\left\| {\bm{q}^{(r)}_{i,T+1} \!-\! \bm{q}^{(r)}_{k,T+1}} \right\|^2} \!\!+\! 2\left({\bm{q}^{(r)}_{i,T+1} \!-\! \bm{q}^{(r)}_{k,T+1}}\right)^T\!\left({\bm{q}_{i,T+1} \!-\! \bm{q}_{k,T+1}}\right), \forall i,k \!\in\! {\cal{D}}, \!i \!\ne\! k,
\end{equation}
where the superscript $(r)$ is the value of the variable at iteration $r$.
Then, the UAV deployment optimization subproblem can be expressed as:
\begin{subequations}\label{optimal_problem3}
\setlength{\abovedisplayskip}{3pt}
\setlength{\belowdisplayskip}{3pt}
\begin{align}\tag{\theequation}
&{\mathop {\min \;}\limits_{{x_{i,T+1}},{y_{i,T+1}},P_{i,T + 1}} \sum\limits_{i \in {\cal{D}}} {{P_{i,T + 1}}} ,}\\
&{\;\;\;\;\;\;\;\;\;\;{\rm{s}}.{\rm{t}}.\;\;\;\;\;\;\;\;\;{P_{i,T + 1}}^{\frac{2}{m+3}} \geqslant {a_j}{d_{ij}^2}, \forall i \in {\cal{D}},\;\forall j \in {{\cal{U}}_i},}\\
&{\;\;\;\;\;\;\;\;\;\;\;\;\;\;\;\;\;\;\;\;\;\;\;-\!{\left\| {\bm{q}^{(r)}_{i,T+1} \!-\! \bm{q}^{(r)}_{k,T+1}} \right\|^2} \!\!+\! 2\left({\bm{q}^{(r)}_{i,T+1} \!-\! \bm{q}^{(r)}_{k,T+1}}\right)^T\!\left({\bm{q}_{i,T+1} \!-\! \bm{q}_{k,T+1}}\right)\!\geqslant\!{d_{\min}}, \;\forall i,k \!\in\! {\cal{D}}, \!i \!\ne\! k,}
\end{align}
\end{subequations}
where $a_j= (\max\left\{lM_j,lN_j\right\})^{\frac{2}{m+3}}$.

Given the user association, problem \eqref{optimal_problem3} is a convex problem due to its convex objective functions and constraints, which can be optimally solved by using the dual method \cite{boyd2004convex}.
The Lagrange function of problem \eqref{optimal_problem3} will be:
\begin{equation}\label{dual}
\begin{split}
\mathcal L=&\sum\limits_{i \in {\cal{D}}} {{P_{i,T + 1}}} +\sum\limits_{i \in {\cal{D}}}\sum_{j\in\mathcal U_i}  \lambda_{ij}^\alpha \left((({x_{i,T+1}} - {v_j}) ^2  +  ({y_{i,T+1}} - {w_j}) ^2  + H^2) a_j   -{P_{i,T + 1}}^{\frac{2}{m+3}}\right) \\
&+\sum\limits_{i \in {\cal{D}}} \sum\limits_{k \in {\cal{D}}, \!k \!\ne\! i}\lambda_{ij}^\beta \left({d_{\min}}+(q^{(r)})^2-2\left(x^{(r)}(x_{i,T+1}-x_{k,T+1})+y^{(r)}(y_{i,T+1}-y_{k,T+1})\right) \right),
\end{split}
\end{equation}
where $(q^{(r)})^2={\left\| {\bm{q}^{(r)}_{i,T+1} \!-\! \bm{q}^{(r)}_{k,T+1}} \right\|^2}$, ${\bm{q}^{(r)}_{i,T+1} \!-\! \bm{q}^{(r)}_{k,T+1}}=(x^{(r)},y^{(r)},0)$, and $\lambda_{ij}^\alpha$ and $\lambda_{ij}^\beta$ are the dual variable associated with constraint (\ref{optimal_problem3}a) and constraint (\ref{optimal_problem3}b), respectively. 

The optimal first-order conditions of (\ref{optimal_problem3}) will be:
\begin{equation}\label{optimal_problem3eq0}
\setlength{\abovedisplayskip}{4pt}
\setlength{\belowdisplayskip}{3pt}
\begin{split}
&\frac{\partial \mathcal L}{\partial P_{i,T + 1}}=1-\frac{2}{m+3}\sum_{j\in\mathcal U_i}  \lambda_{ij}^\alpha {P_{i,T + 1}}^{\frac{-m-1}{m+3}}  =0,
\end{split}
\end{equation}
\begin{equation}\label{optimal_problem3eq1}
\begin{split}
&\frac{\partial \mathcal L}{\partial x_{i,T+1}}= 2\sum_{j\in\mathcal U_i}  \lambda_{ij}^\alpha a_j  ({x_{i,T+1}} - {v_j})- \sum\limits_{k \in {\cal{D}}, \!k \!\ne\! i}\lambda_{ij}^\beta 2x^{(r)}=0,
\end{split}
\end{equation}
\begin{equation}\label{optimal_problem3eq2}
\begin{split}
&\frac{\partial \mathcal L}{\partial y_{i,T+1}} = 2\sum_{j\in\mathcal U_i} \lambda_{ij}^\alpha a_j  ({y_{i,T+1}} - {w_j})- \sum\limits_{k \in {\cal{D}}, \!k \!\ne\! i}\lambda_{ij}^\beta2y^{(r)}=0.
\end{split}
\end{equation}

Solving  \eqref{optimal_problem3eq0} to \eqref{optimal_problem3eq2} yields
\begin{equation}\label{optimal_problem3eq3}
P_{i,T + 1}= {\left(\frac{2}{m+3}\sum_{j\in\mathcal U_i}  \lambda_{ij}^\alpha\right)^{\frac{m+3}{m+1}}},
\end{equation}
\begin{equation}\label{optimal_problem3eq3_2}
x_{i,T+1}=\frac{\sum_{j\in\mathcal U_i}  \lambda_{ij}^\alpha a_j  ({x_{i,T+1}} - {v_j})}{\sum\limits_{k \in {\cal{D}}, \!k \!\ne\! i}\lambda_{ij}^\beta x^{(r)}},\;\;
y_{i,T+1}=\frac{\sum_{j\in\mathcal U_i}  \lambda_{ij}^\alpha a_j  ({y_{i,T+1}} - {w_j})}{\sum\limits_{k \in {\cal{D}}, \!k \!\ne\! i}\lambda_{ij}^\beta y^{(r)}}.
\end{equation}

Given $x_i$, $y_i$, and $P_{i,T + 1}$, the value of $\lambda_{ij}^\alpha$ and $\lambda_{ij}^\beta$ can be determined by the sub-gradient method \cite{bertsekas2009convex}.
The updating procedure is:
\begin{equation}\label{optimal_problem3eq4}
\lambda_{ij}^\alpha= \left[ \lambda_{ij}^\alpha- \gamma \left((({x_{i,T+1}} - {v_j}) ^2  +  ({y_{i,T+1}} - {w_j}) ^2  + H^2) a_j   -{P_{i,T + 1}}^{\frac{2}{m+3}}\right)\right]^+,
\end{equation}
\begin{equation}\label{optimal_problem3eq5}
\lambda_{ij}^\beta= \left[\lambda_{ij}^\beta- \gamma \left({d_{\min}}+(q^{(r)})^2-2\left(x^{(r)}(x_{i,T+1}-x_{k,T+1})+y^{(r)}(y_{i,T+1}-y_{k,T+1})\right) \right)\right]^+,
\end{equation}
where $\gamma$ is a dynamic step size and $[a]^+=\max(a,0)$. 
By solving (\ref{optimal_problem2}), we can obtain the collaborative deployment of multiple UAVs.

With regards to convergence, we have the following theorem:
\begin{theorem}\label{thm5}
The proposed optimization algorithm used to solve UAV deployment problem (\ref{optimal_problem2}) always converges.
\end{theorem}

\itshape \text{Proof:}  \upshape
See Appendix A.
\hfill $\Box$


To find the lower bound of the minimum transmit power, $\inf{P_{i,T + 1}^{\min}}$, we state the following result:  
\begin{proposition}\label{thm1}
If the illumination at the location of user $j$ satisfies the following conditions: 
\begin{equation}
{I_{T+1}^*}({v_j},{w_j}) = \left\{ {\begin{array}{*{20}{l}}
{\!\frac{{{\eta _r} + {n_w}}}{{1 + \sqrt {\frac{{2\pi }}{e}({2^{2{R_j}}} - 1)} }} - {n_w},\;{\eta _r} \geqslant {n_w}\sqrt {\frac{{2\pi }}{e}({2^{2{R_j}}} - 1)},}\\
{\!\;\;\;\;\;\;\;\;\;\;\;\;\;0\;\;\;\;\;\;\;\;\;\;\;\;\;\;\;,\;{\eta _r} < {n_w}\sqrt {\frac{{2\pi }}{e}({2^{2{R_j}}} - 1)},}
\end{array}} \right.
\end{equation}
then the transmit power of each UAV $i$ achieves the lower bound, which is given by: 
\begin{equation}
\inf{P_{i,T+1}^{\min}} = \max_{j\in\mathcal U} \left\{\left( {{\left(n_w+{I_t^*}({v_j},{w_j})\right)}\sqrt {\frac{{2\pi }}{e}({2^{2{R_j}}} - 1)}} \right)l{d_{ij}^{m + 3}}{u_{ij,T+1}}\right\}.
\end{equation}
\end{proposition} 

\itshape \text{Proof:}  \upshape
See Appendix B.
\hfill $\Box$

Proposition \ref{thm1} captures the relationship between the illumination distribution of service area and the minimum transmit power of each UAV.
From Proposition \ref{thm1}, we can see that, given the illuminance requirement $\eta _r$ and data rate constraint $R_j$ of each user $j$, the minimum transmit power of each UAV depends on the illuminance at $({v_j},{w_j})$.
Based on Proposition \ref{thm1}, we can compute the optimal illuminance that allows the transmit power of each UAV $i$ to achieve the lower bound.

\vspace{-0.2cm}
\subsection{User Association and Power Efficiency with Fixed UAV Deployment}
The original optimal problem in (\ref{optimal_problem1}) is combinatorial due to the binary variable $u_{ij,T+1}$. 
Due to the complexity of solving combinatorial problems, the computation is essentially impossible even for a modest-sized wireless network. 
To overcome this, we temporarily adopt the fractional user association relaxation, where association variable $u_{ij,T+1}$ can take on any real value in $[0,1]$. 
We will later show that the optimal solution to $u_{ij,T+1}$ must be either $1$ or $0$ even though the feasible region of $u_{ij,T+1}$ is relaxed to be continuous.
Therefore, the relaxation does not cause any loss of optimality to the final solution of problem (\ref{optimal_problem1}). 
Given the optimal UAV deployment in (\ref{optimal_problem2}), the relaxed problem (\ref{optimal_problem1}) can be formulated as:
\begin{subequations}\label{optimal_problem5}
\begin{align}
\mathop{\min}_{ P_{i,T + 1},{\boldsymbol{u}}_{T+1}} \quad& \sum_{i \in \mathcal D} P_{i,T + 1},  \tag{\theequation}\\
\textrm{s.t.} \qquad
&P_{i,T + 1}\geqslant la_j{d_{ij}^{m+3}}u_{ij,T+1} , \;\forall i\in\mathcal D, \forall j\in\mathcal U,\\
& \sum_{i\in\mathcal D}u_{ij,T + 1} = 1, \;\forall j \in \mathcal U,\\
&  u_{ij,T + 1}\geqslant 0, \;\forall i\in\mathcal D, \forall j\in \mathcal U.
\end{align}
\end{subequations}

To obtain the optimal solution of problem (\ref{optimal_problem5}), we can state the following theorem: 
\begin{theorem}\label{thm2}
For problem (\ref{optimal_problem5}), the optimal user association $u_{ij,T+1}$ and transmit power $P_{i,T+1}$ can be respectively expressed as: 
\begin{equation}\label{PAuser2eq1}
u_{ij,T+1}^*=\left\{ \begin{array}{ll}
\!\!1, &\text{if}\; i =\arg\min_{k\in\mathcal D} \mu_{kj}d_{kj}^{m+3}  \\
\!\!0, &\text{otherwise},
\end{array} \right.
\end{equation}
\begin{equation}\label{PAuser2eq2}
P_{i,T+1}^*=\max_{j\in\mathcal U}la_j{d_{ij}^{m+3}}u_{ij,T+1}^*,
\end{equation}
where $\mu_{ij}$ is the Lagrange multiplier associated with constraint (\ref{optimal_problem5}a),
and $\sum_{j\in\mathcal U} \mu_{ij}\leq 1$.
If there are multiple minimal points in $\arg\min_{k\in\mathcal D}\mu_{kj}d_{kj}^{m+3} $, we will choose any one of them.
\end{theorem}

\itshape \text{Proof:}  \upshape
See Appendix C.
\hfill $\Box$

From Theorem \ref{thm2}, we can see that, even though the feasible region of $u_{ij,T+1}$ is relaxed to be continuous, the optimal solution to problem (\ref{optimal_problem5}) can be effectively solved via its dual problem, while satisfying the discrete constraints ${u_{ij,T + 1}} \in \{ 0,1\},\;\forall i \in {\cal{D}},\forall j \in {{\cal{U}}}$.

The values of $\mu_{ij}$ can be determined by the gradient method \cite{bertsekas2009convex}.
The updating procedure is given by:
\begin{equation}\label{PAuser2eq2_3}
\mu_{ij}=\left[ \mu_{ij}+ \delta (la_j{d_{ij}^{m+3}}u_{ij,T+1} -P_{i,T + 1}) \right]^+,
\end{equation}
where $\delta>0$ is a dynamically chosen step-size sequence.
By iteratively optimizing primal variable and dual variable, the optimal user association and transmit power are obtained.  
Notice that the optimal $u_{ij,T+1}$ is either 0 or 1 according to \eqref{PAuser2eq1}.

\begin{algorithm}[t]
\setlength{\belowcaptionskip}{-1cm}
\scriptsize
\caption{Proposed Algorithm for Deploying UAVs.}
\begin{algorithmic}[1]
\STATE \textbf{Input:} A time series dataset $\boldsymbol{I}$, the set of users' locations, UAV altitude $H$, and the data rate requirement $R_j$.
\STATE \textbf{Initialize:} The user association $\bm{u}_{T+1}$ and dual variables $\bm{\lambda}^{\alpha}$, $\bm{\lambda}^{\beta}$, and $\boldsymbol \mu$.
\STATE Input $\boldsymbol{I}$ into \textbf{Algorithm 1} to predict the illumination distribution ${\boldsymbol{I}_{T+1}}$.
\REPEAT
\STATE Given $\bm{u}_{T+1}$, solve (\ref{optimal_problem2}) using (\ref{optimal_problem3eq3})-(\ref{optimal_problem3eq3_2}).
\STATE Update dual variables  $\bm{\lambda}^{\alpha}$ and $\bm{\lambda}^{\beta}$ using (\ref{optimal_problem3eq4}) and (\ref{optimal_problem3eq5}).
\STATE Given $(x_{i,T+1},y_{i,T+1})$, solve (\ref{optimal_problem5}) using (\ref{PAuser2eq1}) and (\ref{PAuser2eq2}).
\STATE Update dual variable $\boldsymbol \mu$ using (\ref{PAuser2eq2_3}).
\UNTIL the objective value (\ref{optimal_problem}) converges.
\STATE \textbf{Output:} $P = \sum\limits_{i \in {\cal{D}}} {P_{i,T+1}} $.
\end{algorithmic}
\label{algorithm_2}
\end{algorithm}

\vspace{-0.3cm}
\subsection{Complexity and Overhead of the Proposed Algorithms}
The proposed algorithm used to solve problem in (\ref{optimal_problem}) is summarized in \textbf{Algorithm 2}, which includes predicting illumination distribution in service area and iteratively optimizing UAV deployment, user association, and energy efficiency.
The complexity of the proposed algorithm lies in training an integrated GRU and CNN predictive model and iteratively updating UAV location $(x_i,y_i)$ and user association ${\bm{u}}_{T+1}$.
The complexity for training an integrated GRU and CNN predictive model is detailed in the following lemmas: 
\begin{lemma}\label{thm3}
For the CNN-based illumination distribution encoder and DeCNN-based decoder, the complexity are both $\mathcal O(\sum\limits_{l = 1}^L {{\lambda_l}^2{S^2}{K_c^{l}}{K_c^{l-1}}})$.
\end{lemma}

\itshape \text{Proof:}  \upshape
See Appendix D.
\hfill $\Box$

From Lemma \ref{thm3}, we can see that the complexity of CNN encoder and DeCNN decoder depends on the size and number of feature maps in each layer. 

\begin{lemma}\label{thm4}
For the GRU-based illumination distribution predictor, the complexity is given as ${\mathcal O}\left( {T{D_h}(N + {D_h})} \right)$. 
\end{lemma}

\itshape \text{Proof:}  \upshape
See Appendix E.
\hfill $\Box$

From Lemma \ref{thm4}, we can see that the complexity of GRU predictor depends on the length of input time series and the size of weight matrices.
Since the integrated GRU and CNN predictive model is trained by the BS which has enough computational ability for training, the overhead of training the predictive model can be ignored. 
Meanwhile, once the training process is completed, the trained integrated GRU and CNN model can be used to predict the illumination distribution in a long term period.

Next, we investigate the complexity of solving the optimization problem, which lies in solving two subproblems: UAV deployment problem and user association problem.
For the UAV deployment problem, the overhead of calculating $(x_i,y_i)$ of each UAV $i$ in \eqref{optimal_problem3eq3_2} is $\mathcal O\left(L_1D^2U\right)$, where $L_1$ is the average number of iterations until (\ref{optimal_problem2}) convergence.
For the user association problem, the overhead of obtaining $\boldsymbol u_{T+1}$ from \eqref{PAuser2eq1} is $\mathcal O(L_2DU)$, where $L_2$ is the average iteration number until (\ref{optimal_problem5}) converges.
The two subproblems are solved by dual method.
According to \cite{bertsekas2009convex}, a sharp estimate of $L_1$ and each $L_2$ can be expressed as $\mathcal O \left(\frac{1}{\sqrt{\epsilon}}\right)$, where $\epsilon$ is the accuracy of the dual method.
As a result, the complexity to solve the UAV deployment problem and the user association problem can be further simplified as $\mathcal O \left( \frac{D^2U}{\sqrt{\epsilon}}\right)$ and $\mathcal O \left( \frac{{DU}}{\sqrt{\epsilon}}\right)$, respectively.
Therefore, the UAVs only need to implement the optimization algorithm with the linear complexity and thus reducing energy consumption of UAVs.

\vspace{-0.3cm}
\section{Simulation Results and Analysis}
\label{sec:5}
For our simulations, an $80$~m $\times$ $80$~m square area is considered with $U=40$ uniformly distributed users and $D=4$ UAVs deployed at a height of $20$~m. 
The downlink rate requirement $R_j$ of each user $j$ is generated randomly and uniformly over [0.5,1.5] Mbps. 
Other parameters are listed in Table \uppercase\expandafter{\romannumeral1}. 
Furthermore, the homogeneous regularized LoS probability $\bar{B}$ in (\ref{power}) is set as the value corresponding to the elevation angle of $90^\circ$.
The time series illumination data used to train integrated GRU and CNN predictive model is a dataset of average radiance composite nighttime remote sensing images, obtained from the Earth observations group (EOG) at NOAA/NCEI \cite{EOG}.
Since there is no public illumination dataset collected by UAVs, we use the satellite remote sensing dataset to verify the performance of our proposed prediction and optimization algorithm.

\begin{table}\scriptsize
\vspace{-0.3cm}
\setlength{\belowcaptionskip}{-10pt}
\setlength{\abovedisplayskip}{-15pt}
\newcommand{\tabincell}[2]{\begin{tabular}{@{}#1@{}}#1.6\end{tabular}}
\renewcommand\arraystretch{1}
\caption[table]{{System Parameters}}
\centering
\begin{tabular}{|c|c|c|c|c|c|}
\hline
\!\textbf{Parameters}\! \!\!& \textbf{Value} &\! \textbf{Parameters} \!& \textbf{Value} &\! \textbf{Parameters} \!& \textbf{Value} \\
\hline
$\Phi $ & $90^\circ$ &  $\Psi_c$ & $90^\circ$ & $\rho $ & 0.5 m$^2$\\
\hline
$\xi$& 0.8 Amp./W & $n_e$ & 1.5 & $n _w$ & $1 \times {10^{ - 10}}$\\
\hline
$X$ & 10 &$Y$& 0.6 & $\eta _r$ & $5 \times {10^{ - 4}}$\\
\hline
$S$&3 & $N$& 256 & $D_h$ & $64$\\
\hline
$D_q$ & 16  & $L$& 4 & $\gamma$& 0.01\\
 \hline
 $\delta$& 0.01 & $e$& $10^{4}$ & $\epsilon$& $10^{-4}$ \\
\hline
\end{tabular}
\end{table}
%
\begin{figure}[t]
\centering
\setlength{\abovecaptionskip}{-0cm}
\setlength{\belowcaptionskip}{-0.8cm}
\includegraphics[width=15cm]{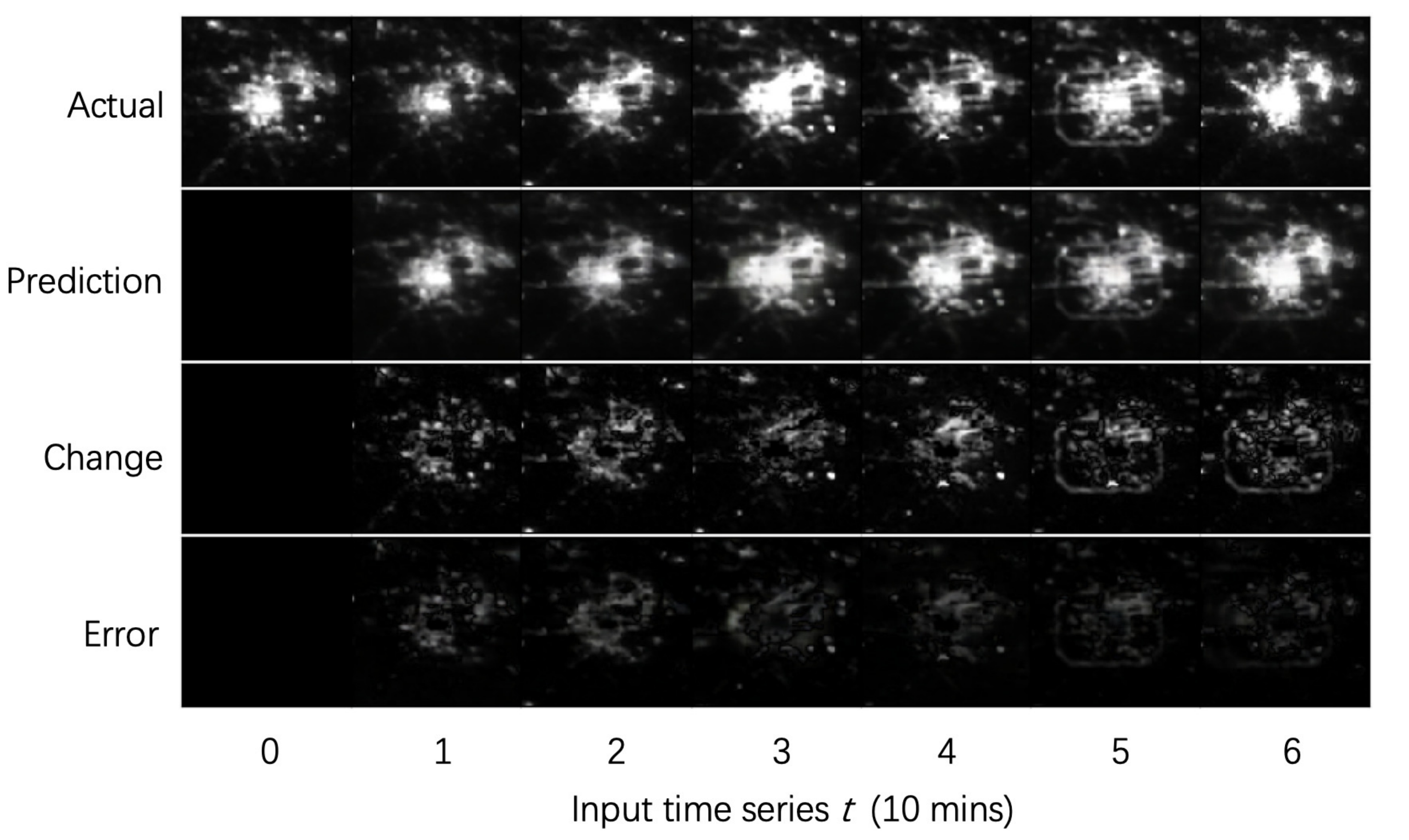}
\caption{Predicted illumination distribution of the target area.}
\label{fig2}
\end{figure}
Fig. \ref{fig2} shows how the predicted illumination distributions change as the input time series change. 
We randomly select an area for the predictions of illumination distribution. 
In Fig. \ref{fig2}, we can see that the prediction at the first time step is initialized to zero.
Fig. \ref{fig2} also shows that, as time elapses, the accuracy of illumination distribution prediction generated by the model increases.
This is because the proposed model can build a relationship between the prediction and the historical illumination distribution.
As the number of input historical illumination distribution increases, the proposed model can extract obtain more time-varying information about the illumination distribution.

\begin{figure}[t]
\centering
\includegraphics[width=16.5cm]{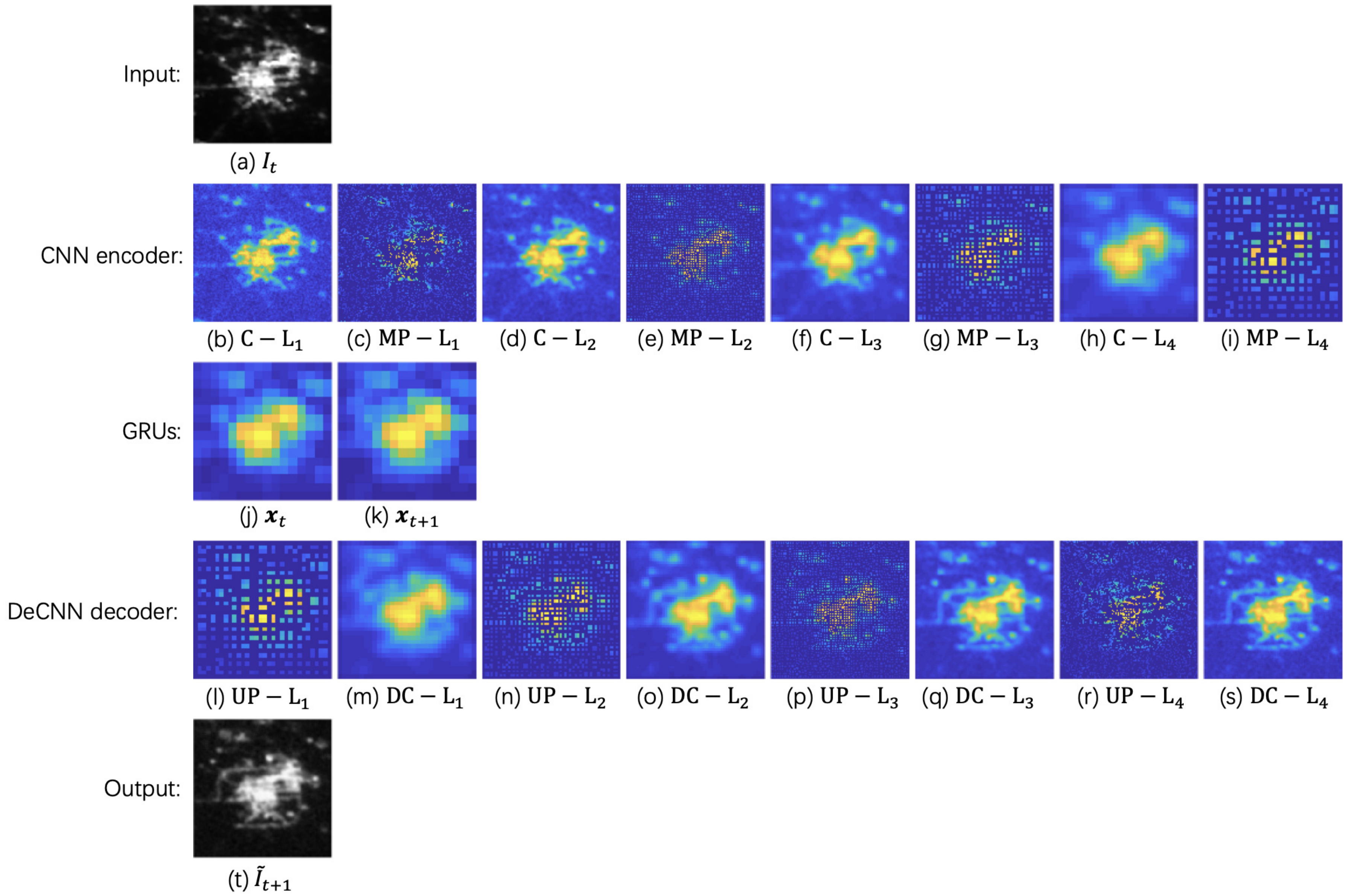}
\caption{Visualization of extracted features in the proposed predictive model.}
\vspace{-0.9cm}
\label{fig9}
\end{figure}

In Fig.~\ref{fig9}, we show how the integrated GRU and CNN model predicts the illumination distribution at next time slot.
Here, we combine the representative features in each layer for effective visualization.
Fig.~\ref{fig9}(a) is an actual illumination distribution at time slot $t$ and it is also an input of the proposed predictive model.
Figs.~\ref{fig9}(b) to \ref{fig9}(j) show the extracted feature maps in the CNN encoding components, which are extracted from $256 \times 256$ convolutional layer, $128 \times 128$ max-pooling layer, $128 \times 128$ convolutional layer, $64 \times 64$ max-pooling layer, $64 \times 64$ convolutional layer, $32 \times 32$ max-pooling layer, $32 \times 32$ convolutional layer, $16 \times 16$ max-pooling layer, and $16 \times 16$ flatten layer, respectively.
Fig.~\ref{fig9}(k) visualizes the predicted features of illumination distribution at time slot $t+1$, $\bm{x}_{t+1}$, obtained by GRUs.
Based on $\bm{x}_{t+1}$, Figs.~\ref{fig9}(l) to \ref{fig9}(s) are the output maps in the DeCNN decoding components, which are reconstructed from $32 \times 32$ unpooling layer, $32 \times 32$ deconvolutional layer, $64 \times 64$ unpooling layer, $64 \times 64$ deconvolutional layer, $128 \times 128$ unpooling layer, $128 \times 128$ deconvolutional layer, $256 \times 256$ unpooling layer, and $256 \times 256$ deconvolutional layer, respectively.
Fig.~\ref{fig9}(t) shows the predicted illumination distribution at time slot $t+1$ output from the integrated GRU and CNN model.
From Figs.~\ref{fig9}(b) to \ref{fig9}(j) we can see that the CNN encoder captures the boundary information and shading information of the illumination distribution.
This is because the features that are closely related to the change of illumination distribution are amplified through forward-propagation while noisy features from background are suppressed. 
From Figs.~\ref{fig9}(l) to \ref{fig9}(s) we can see that the coarse-to-fine structures of the illumination distribution are reconstructed after the predicted features propagate through DeCNN decoder layers.
This is due to the fact that, unpooling layers trace predicted features back to the original locations in service area and deconvolutional layers effectively reconstruct the detailed structure of illumination distribution based on the learned weights.

\begin{figure}[t]
\centering
\setlength{\abovecaptionskip}{-0cm}
\setlength{\belowcaptionskip}{-0.8cm}
\includegraphics[width=11cm]{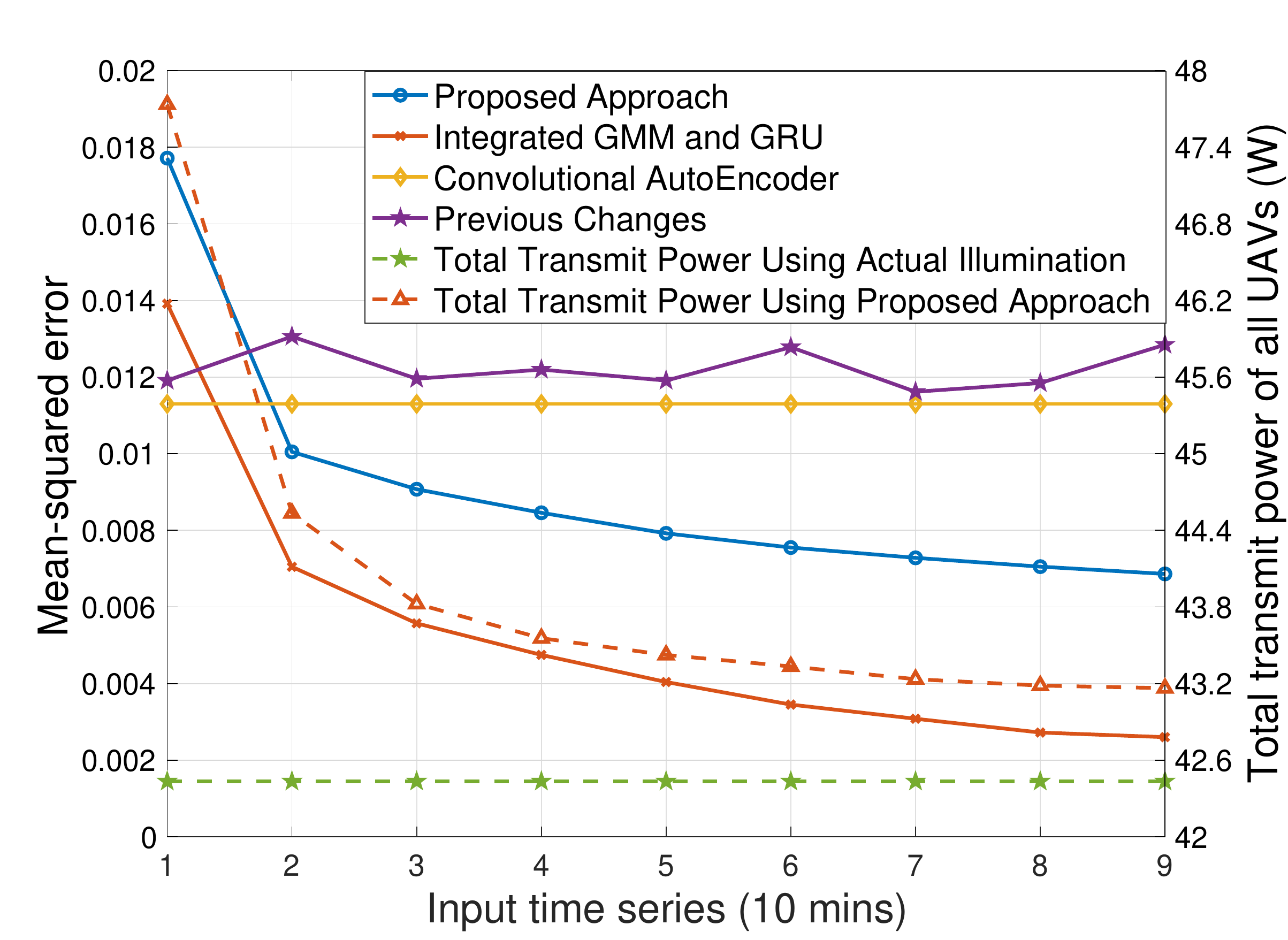}
\caption{Prediction accuracy of the illumination distribution and the required sum power of UAVs as a function of the size of input series.}
\label{fig3}
\end{figure}

\begin{figure}[t]
\centering
\setlength{\abovecaptionskip}{-0cm}
\setlength{\belowcaptionskip}{-0.8cm}
\includegraphics[width=11cm]{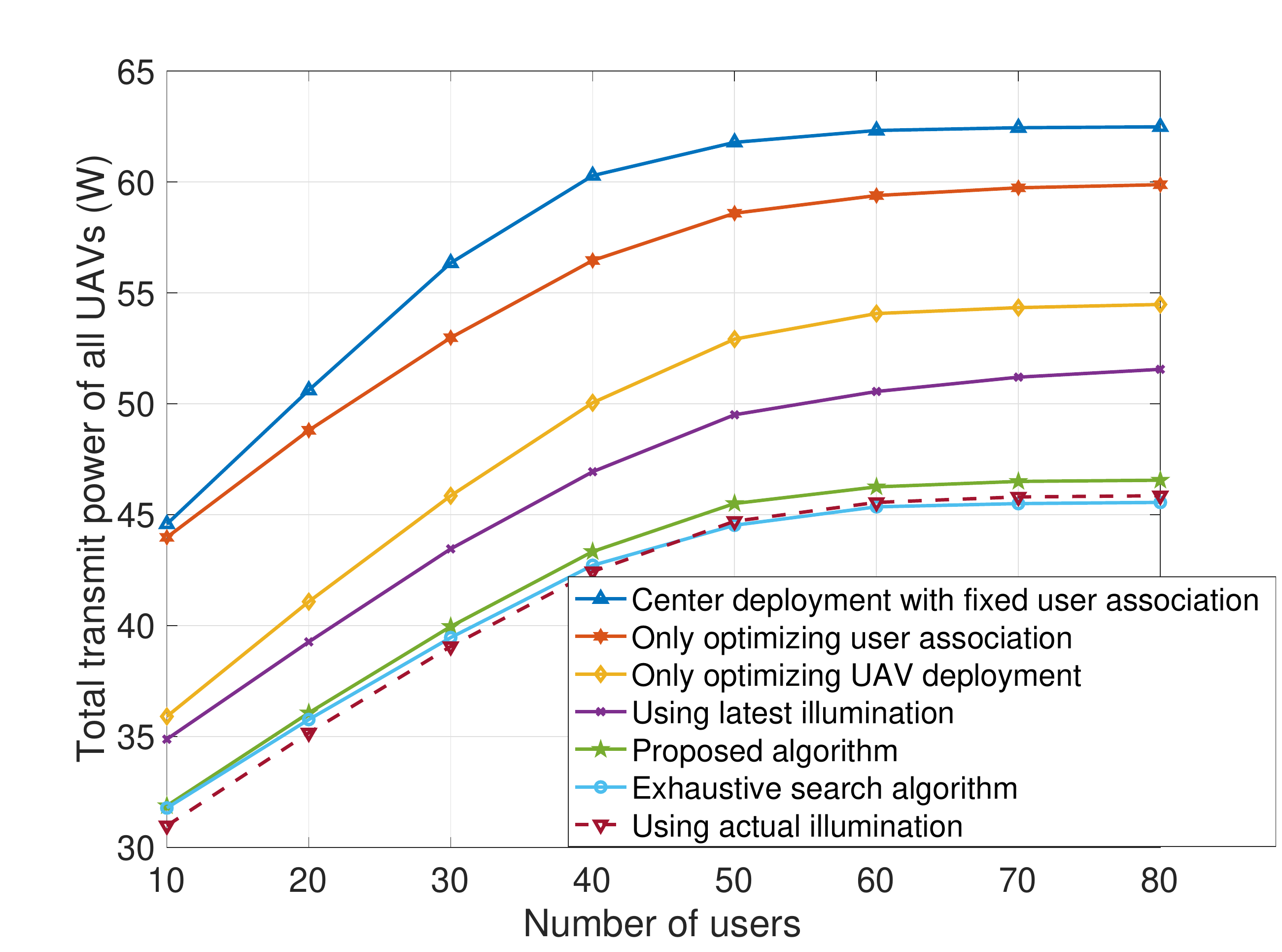}
\caption{The required sum power of UAVs as the number of users varies.}
\vspace{-0.1cm}
\label{fig4}
\end{figure}

Fig. \ref{fig3} shows how the prediction accuracy of the illumination distribution on two test service areas changes as the size of input time series $t$ varies.
In Fig. \ref{fig3}, for comparison, we include the results of an integrated GMM and GRU model \cite{wyy2019GC} and an autoencoder in \cite{autoencoder} trained on single time interval illumination distribution.
210 area samples are used to train the proposed model, with each area containing 78 illumination distributions in time series.
We randomly choose 5\% of each illumination series for validation and testing, and discard the chosen continuous segments from the training set.
From Fig. \ref{fig3}, we can see that, as the length of input illumination series $t$ increases, the mean-square error (MSE) of the proposed model decreases, while the variation of the illumination distribution over each time slot is random.
This is due to the fact that, as the input series $t$ increases, the proposed model can accumulate information on the change of illumination distribution.
The average MSE of training data prediction and test data prediction are $6.01 \times {10^{ - 4}}$ and $6.03 \times {10^{ - 4}}$, respectively. 
Fig. \ref{fig3} also shows that the proposed model can yield up to 46.5\% and 53.6\% reduction in terms of MSE compared with integrated Gaussian mixture model (GMM) and GRU model and conventional autoencoder, respectively.
These gains stem from the fact that, the proposed model can simultaneously extract the spatial and temporal features of historical illumination distributions so as to accurately predict future illumination distributions.
In Fig. 5, the MSE for conventional autoencoder is a constant due to the fact that the autoencoder can only analyze the input data over two consecutive time slots. 
Therefore, even if the time series of input illumination distribution becomes longer, the autoencoder still generates the illumination distribution prediction based on the single input at last time slot.
Fig. \ref{fig3} also shows that, as the length of input illumination series $t$ increases, the gap between the minimum transmit power resulting from the proposed algorithm and the minimum transmit power resulting from the optimization algorithm using the actual illumination distribution decreases.
This is because, as the input series $t$ increases, the accuracy of illumination distribution prediction of the proposed algorithm improves and the optimal UAV deployment and user association is related to the illumination distribution.
Hence, the accurate illumination distribution predictions can effectively reduce the required transmit power of UAVs.

\begin{figure}[t]
\centering
\setlength{\abovecaptionskip}{-0cm}
\setlength{\belowcaptionskip}{-0.7cm}
\includegraphics[width=11cm]{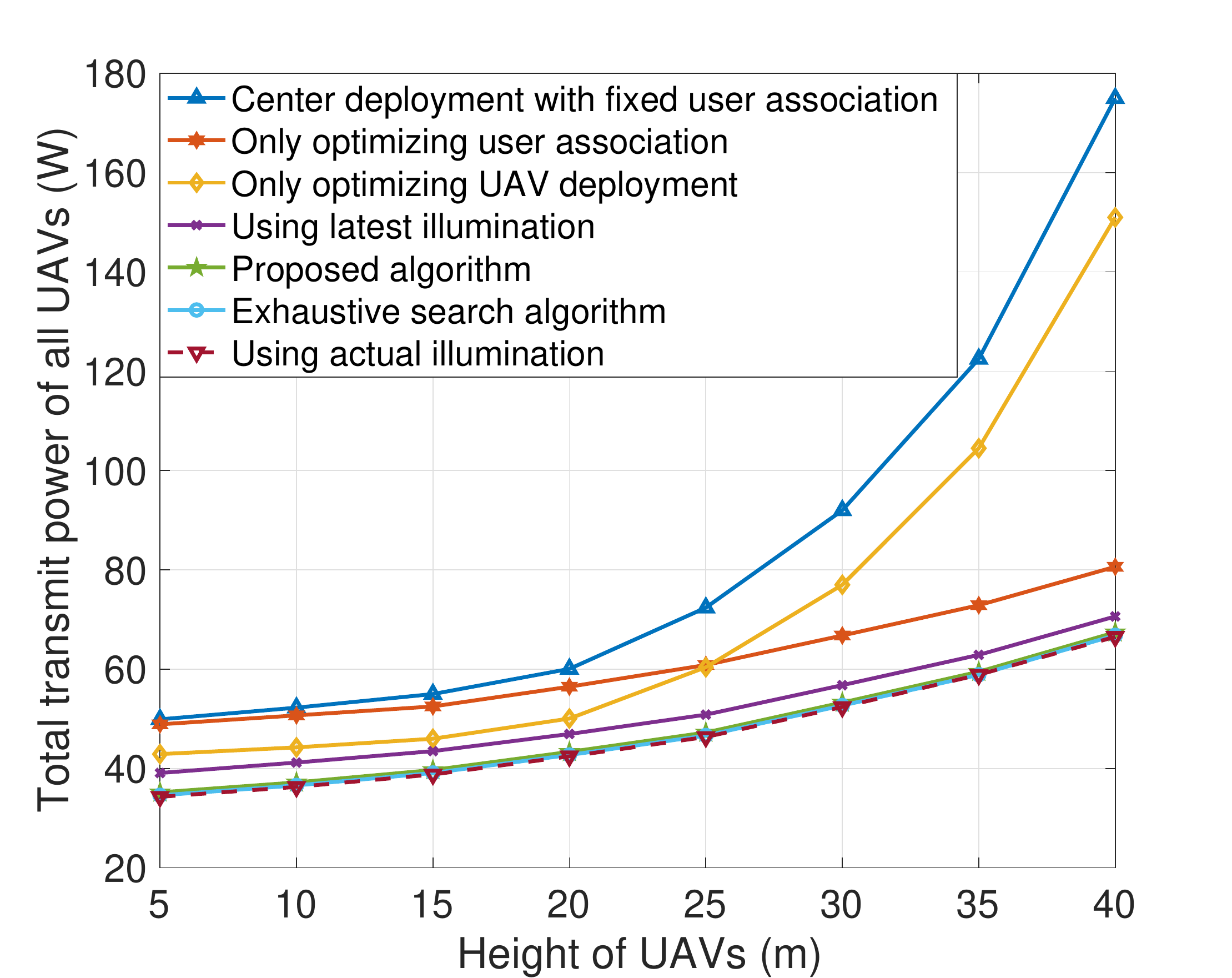}
\caption{The required sum power of UAVs as the height of UAVs varies.}
\vspace{-0.1cm}
\label{fig7}
\end{figure}

Fig. \ref{fig4} shows how the transmit power used to meet the users' data rate and illumination requirements changes as the number of users varies. 
In Fig.~\ref{fig4}, we can see that the proposed algorithm can reduce transmit power by up to 37.4\% on average compared to a conventional optimal UAV deployment which is the center of the service area. 
In addition, the proposed algorithm can yield up to 31.3\%, and 15.5\% reductions in terms of the total transmit power compared to only optimizing the user association optimization and only optimizing the UAV deployment, respectively.
These reductions are due to the fact that the power required by the users is related to the illumination of the service area and the deployment of the associated UAV. 
The proposed algorithm can iteratively optimize user association and UAV deployments, which will reduce the total transmit power of all the UAVs.
From Fig.~\ref{fig4} we can also see that the proposed algorithm can reduce up to 9.3\% transmit power compared to an algorithm that uses the latest illumination distribution.
Moreover, the proposed algorithm is closer to the UAV deployment optimization using actual illumination distribution and the gap between the two schemes is less than 2.8\%. 
This is because the proposed prediction algorithm can accurately predict the illumination distribution so as to optimize UAV deployment. 
In Fig.~\ref{fig4}, we can also see that the gap between the proposed iterative algorithm and the exhaustive search algorithm is less than 1.5\%, which indicates that the proposed iterative algorithm approaches the near globally optimal solution.
Fig.~\ref{fig4} also shows that, as the number of users increases, the performance gain of the proposed deployment becomes less significant. 
This is because when enough users are considered, the users will be uniformly distributed in the square and the optimal position of the UAV will be fixed.

\begin{figure}[t]
\centering
\setlength{\abovecaptionskip}{-0cm}
\setlength{\belowcaptionskip}{-0.8cm}
\includegraphics[width=11cm]{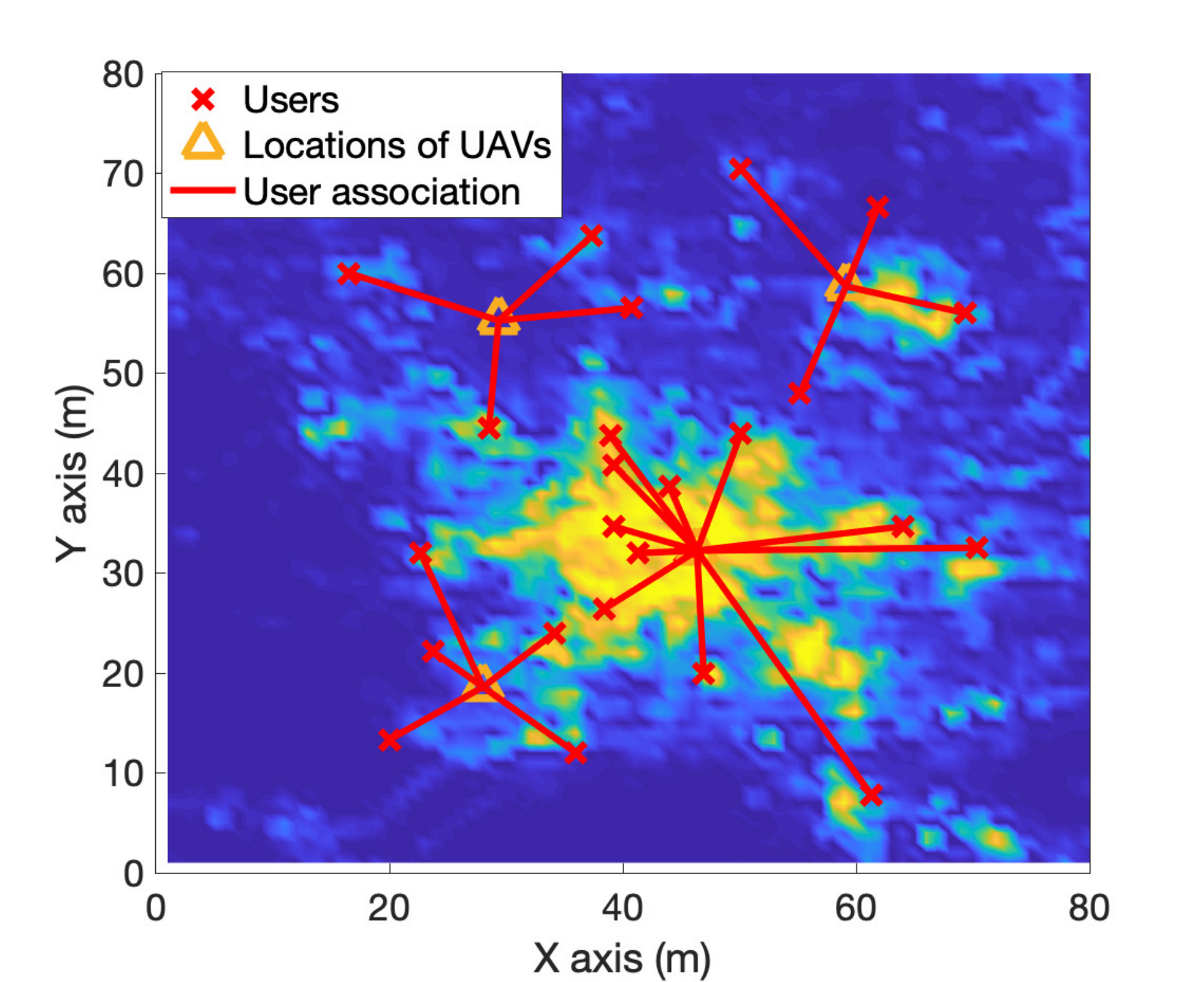}
\caption{Optimal deployment of UAVs using the proposed algorithm.}
\vspace{-0.1cm}
\label{fig5}
\end{figure}

Fig. \ref{fig7} shows how the transmit power used to meet the users' data rate and illumination requirements changes as the height of UAVs varies. 
In Fig.~\ref{fig7}, we can see that, as the height of the UAVs increases, the total transmit power of all algorithms increases since the deployment of UAVs at a high altitude increases the distance from the user to the associated UAV.
In Fig.~\ref{fig7}, we can also see that the proposed algorithm achieves up to 68.9\% gain in terms of transmit power reduction compared to a conventional optimal UAV deployment without considering the illumination distribution and user association. 
Fig.~\ref{fig7} also shows that the proposed algorithm can yield up to 29.2\%, and 43.0\% reductions in terms of the total transmit power compared to only optimizing the user association optimization and only optimizing the UAV deployment, respectively.
This implies that, as the height of the UAVs increases, the transmit power gain achieved by optimizing the user association becomes more significant than the gain achieved by optimizing the UAV deployment.
This is because, when the UAVs are deployed at a very high altitude, the proposed algorithm prefers to associate all the users with as few UAVs as possible, while other UAVs are idle.
Therefore, the optimal user association obtained by the proposed algorithm will significantly limit the increase in total transmit power of all the UAVs cause by the long distance between UAVs and users.

In Fig. \ref{fig5}, we show an example of how the proposed algorithm can optimize the deployment of UAVs. 
In the example, four UAVs are deployed at a height of $20$~m to serve a $80$~m $\times$ $80$~m square area.
Fig. \ref{fig5} shows the optimal location of each UAV and the optimal user association which result in the minimum total transmit power.
From Fig. \ref{fig5}, we can see that users in the area with strong illumination are served by a single UAV and scattered users are served by other UAVs.
This is because, the minimum transmit power of each UAV depends on the maximum requirement of its associated users, which usually occurs at the bright area.
Once the users located in the bright area are simultaneously satisfied by one UAV, the minimum transmit power of other UAVs reduce significantly, thus achieving a minimum total transmit power.


\vspace{-0.2cm}
\section{Conclusion}
\label{sec:6}
In this paper, we have developed a novel UAV deployment framework for dynamically optimizing the locations and user association of UAVs in a VLC-enabled UAV based network. 
We have formulated an optimization problem that seeks to minimize the transmit power while meeting the illumination and communication requirements of each user. 
To solve this problem, we have developed an integrated GRU and CNN prediction algorithm, which can model the long-term historical illumination distribution and predict the future illumination distribution. 
We have then transformed the nonconvex original problem into convex reformulation through physical relaxation of the user association.
Therefore, the optimal solution of the optimization problem is obtained by an iterative algorithm.
Simulation results have shown that the proposed approach yields significant power reduction compared to conventional approaches.

\vspace{-0.2cm}
\section*{Appendix}

\vspace{-0.3cm}
\subsection{Proof of Theorem 1}
To prove the convergence of the algorithm used to solve problem (\ref{optimal_problem2}), we first need to prove that the solution of problem (\ref{optimal_problem3}) at each iteration is optimal.
At each iteration, problem (\ref{optimal_problem3}) is convex and it is solved by the dual method using (\ref{dual})-(\ref{optimal_problem3eq5}).
According to \cite{bertsekas2009convex}, the dual method can find an optimal solution for problem (\ref{optimal_problem3}) at each iteration.
To prove that the proposed algorithm used to solve the entire problem (\ref{optimal_problem2}) converges, we assume that:
\begin{equation}
f(\bm{q}_{i,T+1},\bm{q}_{k,T+1}) ={\left\| {\bm{q}_{i,T+1} - \bm{q}_{k,T+1}} \right\|^2}
\end{equation}
and
\begin{equation}
g^{(r)}(\bm{q}_{i,T+1}, \bm{q}_{k,T+1})=-{\left\| {\bm{q}^{(r)}_{i,T+1} - \bm{q}^{(r)}_{k,T+1}} \right\|^2} + 2\left({\bm{q}^{(r)}_{i,T+1} - \bm{q}^{(r)}_{k,T+1}}\right)^T\left({\bm{q}_{i,T+1}-\bm{q}_{k,T+1}}\right).
\end{equation}
Then, (\ref{optimal_problem2}c) and (\ref{optimal_problem3}b) at iteration $r$ can be represented as $f\;(\;\bm{q}_{i,T+1}\;,\;\bm{q}_{k,T+1}\;)\;\geqslant\; d_{\min}$ and \\
$g^{(r)}(\bm{q}_{i,T+1}, \bm{q}_{k,T+1})\geqslant d_{\min}$, respectively.
Since (\ref{optimal_problem3}b) is the first-order Taylor expansion of (\ref{optimal_problem2}c), we have: 
\begin{equation}\label{P1}
g^{(r)}\left(\bm{q}_{i,T+1}, \bm{q}_{k,T+1}\right)\geqslant f\left(\bm{q}_{i,T+1},\bm{q}_{k,T+1}\right), \; \forall \bm{q}_{i,T+1}, \forall \bm{q}_{k,T+1},
\end{equation}
\begin{equation}\label{P2}
g^{(r)}\left(\bm{q}^{(r-1)*}_{i,T+1}, \bm{q}^{(r-1)*}_{k,T+1}\right)= f\left(\bm{q}^{(r-1)*}_{i,T+1},\bm{q}^{(r-1)*}_{k,T+1}\right),
\end{equation}
where $\bm{q}^{(r-1)*}_{i,T+1}$ and $\bm{q}^{(r-1)*}_{k,T+1}$ are the optimal solution of problem (\ref{optimal_problem3}) at iteration $r-1$. 
Given (38) and (39), we can conclude that, every $\bm{q}^{(r-1)*}_{i,T+1}$ and $\bm{q}^{(r-1)*}_{k,T+1}$ satisfy all conditions of the original problem (\ref{optimal_problem2}).
Let $\bm{q}^{(r)}_{i,T+1}=\bm{q}^{(r-1)*}_{i,T+1}$ and $\bm{q}^{(r)}_{k,T+1}=\bm{q}^{(r-1)*}_{k,T+1}$ at each iteration $r$.
Substituting $\bm{q}^{(r)}_{i,T+1}$ and $\bm{q}^{(r)}_{k,T+1}$ into (\ref{optimal_problem2}), we can calculate the total transmit power, which is represented by $\sum\limits_{i \in {\cal{D}}}{P^{(r)}_{i,T+1}}$.
Since $\bm{q}^{(r)*}_{i,T+1}$ and $\bm{q}^{(r)*}_{k,T+1}$ are the optimal solutions for problem (\ref{optimal_problem3}) using the start point $\bm{q}^{(r-1)*}_{i,T+1}$ and $\bm{q}^{(r-1)*}_{k,T+1}$, we have $\sum\limits_{i \in {\cal{D}}} {P^{(r)*}_{i,T+1}} \leqslant \sum\limits_{i \in {\cal{D}}}{P^{(r-1)*}_{i,T+1}}$.
Since the total transmit power $\sum\limits_{i \in {\cal{D}}}{P_{i,T+1}}$ is finitely lower-bounded by zero, the optimization algorithm for solving (\ref{optimal_problem2}) must converge.
This completes the proof.

\vspace{-0.2cm}
\subsection{Proof of Proposition 1}
\label{Proposition_1}
Based on (\ref{power}) and (\ref{optimal_problem3}), the minimum transmit power of UAV $i$ to satisfy the requirements of user $j$ can be given by:
\vspace{-0.15cm}
\begin{equation}\label{power_res}
\setlength{\abovedisplayskip}{2pt}
\setlength{\belowdisplayskip}{2pt}
{P_{ij,T+1}^{\min}} =\max \left\{M_j, N_j \right\} ld_{ij}^{m+3}, \forall j \in {{\cal{U}}_i}.
\end{equation}
where $M_j={{\eta _r} - {I_{t}(v_j,w_j)}}$ and $N_j={({n_w} + {I_{t}(v_j,w_j)})\sqrt {\frac{{2\pi }}{e}({2^{2{R_j}}} - 1)}}$.
Given illuminance and data rate requirements of each user $j$, to obtain the lower bound of ${P_{ij,T+1}^{\min}}$, we derive the first derivative with respect ${I_{T+1}(v_j,w_j)}$ as:
\vspace{-0.2cm} 
\begin{equation}\label{first_derivative}
\frac{{\partial {P_{ij,T+1}^{\min}}}}{{\partial {I_{T+1}}({v_j},{w_j})}} = \left\{ {\begin{array}{*{20}{l}}
{\! \;\;\;\;\;\;\;\;\;\; -ld_{ij}^{m+3}\;\;\;\;\;\;\;\;\;\; ,\; M_j > N_j,}\\
{\! {\sqrt {\frac{{2\pi }}{e}({2^{2{R_j}}} - 1)}}ld_{ij}^{m+3}\;,\;M_j < N_j,}
\end{array}}
\right.
\end{equation}
Since $-ld_{ij}^{m+3} < 0$ and ${\sqrt {\frac{{2\pi }}{e}({2^{2{R_j}}} - 1)}}ld_{ij}^{m+3}  > 0$, there is a unique ${I_{T+1}}({v_j},{w_j})$ that allows the minimum transmit power to reach the lower bound.
To find the lower bound of the minimum transmit power of UAV $i$, we need to compare the power used to satisfy the illumination requirement of its associated users and the transmit power used to satisfy the data rate requirement of its associated users.
Next, we analyze the optimal illumination, $I_{T+1}^*(v_j,w_j)$, that allows ${P_{ij,T+1}^{\min}}$ to reach the lower bound.

If $M_j < N_j$ for $\forall I_{T+1}(v_j,w_j) \geqslant 0$, that is ${\eta _r}<{{n_w}\sqrt {\frac{{2\pi }}{e}({2^{2{R_j}}} - 1)}}$, we have ${P_{ij,{T+1}}^{\min}} = N_j ld_{ij}^{m+3}$.
Since $\frac{{\partial {P_{ij,T+1}^{\min}}}}{{\partial {I_{T+1}}({v_j},{w_j})}}  ={\sqrt {\frac{{2\pi }}{e}({2^{2{R_j}}} - 1)}}ld_{ij}^{m+3}> 0$ and $I_{T+1}(v_j,w_j) \geqslant 0$, the optimal $I_{T+1}^*(v_j,w_j)$ that allows ${P_{ij,T+1}^{\min}}$ to reach the lower bound will be:
\vspace{-0.15cm}
\begin{equation}
\setlength{\abovedisplayskip}{2pt}
\setlength{\belowdisplayskip}{2pt}
I_{T+1}^*(v_j,w_j)=0,
\end{equation}
and the lower bound of the minimum transmit power of UAV $i$ to satisfy its associated user $j$ will be:
\vspace{-0.15cm}
\begin{equation}\label{min1}
\inf{P_{ij,{T+1}}^{\min}} = {{n_w}\sqrt {\frac{{2\pi }}{e}({2^{2{R_j}}} - 1)}} l d_{ij}^{m+3}.
\end{equation}

Otherwise, we have ${\eta _r} \geqslant {{n_w}\sqrt {\frac{{2\pi }}{e}({2^{2{R_j}}} - 1)}}$. From (\ref{first_derivative}), we can see that ${P_{ij,T+1}^{\min}}$ achieves the minimum value when $M_j=N_j$, that is ${\eta _r} - {I_{T+1}^*}({v_j},{w_j}) = ({n_w} + {I_{T+1}^*}({v_j},{w_j}))\sqrt {\frac{{2\pi }}{e}({2^{2{R_j}}} - 1)}$. 
Then, we have ${\eta _r} - {n_w}\sqrt {\frac{{2\pi }}{e}({2^{2{R_j}}} - 1)} = {I_{T+1}^*}({v_j},{w_j})\left(\sqrt {\frac{{2\pi }}{e}({2^{2{R_j}}} - 1)}  + 1\right)$.
Therefore, the optimal $I_{T+1}^*(v_j,w_j)$ will be: 
\vspace{-0.15cm}
\begin{equation}
\begin{aligned}
{I_{T+1}^*}({v_j},{w_j}) &= \frac{{{\eta _r} - {n_w}\sqrt {\frac{{2\pi }}{e}({2^{2{R_j}}} - 1)} }}{{1 + \sqrt {\frac{{2\pi }}{e}({2^{2{R_j}}} - 1)} }}\\
&=\frac{{{\eta _r} + {n_w}}}{{1 + \sqrt {\frac{{2\pi }}{e}({2^{2{R_j}}} - 1)} }} - {n_w},
\end{aligned}
\end{equation}
and the lower bound of the minimum transmit power of UAV $i$ to satisfy its associated user $j$ will be:
\vspace{-0.15cm}
\begin{equation}\label{min2}
\inf{P_{ij,T+1}^{\min}} = {{\left(n_w+{I_{T+1}^*}({v_j},{w_j})\right)}\sqrt {\frac{{2\pi }}{e}({2^{2{R_j}}} - 1)}} l d_{ij}^{m+3}.
\end{equation}

Therefore, the optimal illumination at the location of user $j$ is given by:
\begin{equation}\label{op_illu}
\setlength{\abovedisplayskip}{2pt}
\setlength{\belowdisplayskip}{2pt}
{I_{T+1}^*}({v_j},{w_j}) = \left\{ {\begin{array}{*{20}{l}}
{\!\frac{{{\eta _r} + {n_w}}}{{1 + \sqrt {\frac{{2\pi }}{e}({2^{2{R_j}}} - 1)} }} - {n_w},\;{\eta _r} \geqslant {n_w}\sqrt {\frac{{2\pi }}{e}({2^{2{R_j}}} - 1)},}\\
{\!\;\;\;\;\;\;\;\;\;\;\;\;\;0\;\;\;\;\;\;\;\;\;\;\;\;\;\;\;,\;{\eta _r} < {n_w}\sqrt {\frac{{2\pi }}{e}({2^{2{R_j}}} - 1)},}
\end{array}} \right.
\end{equation}
Based on (\ref{min1}) and (\ref{min2}), the lower bound of the minimum transmit power of each UAV $i$ at time $T+1$ is given as:
\vspace{-0.15cm}
\begin{equation}
\inf{P_{i,T+1}^{\min}} = \max_{j\in\mathcal U} \left\{\left( {{\left(n_w+{I_{T+1}^*}({v_j},{w_j})\right)}\sqrt {\frac{{2\pi }}{e}({2^{2{R_j}}} - 1)}} \right)l{d_{ij}^{m + 3}}{u_{ij,T+1}}\right\}.
\end{equation}
This completes the proof.

\vspace{-0.3cm}
\subsection{Proof of Theorem 2}
The dual problem of problem (\ref{optimal_problem5}) with relaxed constraints can be given by:
\vspace{-0.15cm}
\begin{equation}\label{dueq1}
\setlength{\abovedisplayskip}{2pt}
\setlength{\belowdisplayskip}{2pt}
\mathop{\max}_{\boldsymbol \beta}\quad D(\boldsymbol \beta),
\end{equation}
where
\vspace{-0.1cm}
\begin{equation}\label{dueq1_2}
D( \boldsymbol \beta)=\left\{ \begin{array}{ll}
\!\!\!\mathop{\min}\limits_{P_{i,T+1}, \boldsymbol u_{T+1}} 
&  \mathcal L (P_{i,T+1}, \boldsymbol u_{T+1},  \boldsymbol \beta)
\\
 \quad \rm{s.t.}& \sum_{i\in\mathcal D}u_{ij,T + 1} = 1, \quad \forall j \in \mathcal \mathcal U,\\
&  u_{ij,T + 1}\geqslant 0,\quad \forall i\in\mathcal D, \forall j\in \mathcal U,
\end{array} \right.
\end{equation}
with
\vspace{-0.2cm}
\begin{align}\label{dueq2}
\mathcal L (P_{i,T+1}, \boldsymbol u_{T+1},  \boldsymbol \beta)=& \sum_{i \in \mathcal D} P_{i,T + 1} +\sum_{i \in \mathcal D}\sum_{j \in \mathcal U}\beta_{ij}(la_j{d_{ij}^{m+3}}u_{ij,T+1} -P_{i,T + 1})
\end{align}
and $\pmb \beta=\{\beta_{ij}\}$.

To minimize the objective function in (\ref{dueq1}), which is a linear combination of $u_{ij,T+1}$, we should let the smallest association coefficient corresponding to the $u_{ij,T+1}$ be 1 among all UAV $i$ with given user $j$.
Therefore, the optimal $u_{ij,T+1}^*$ is thus given as: 
\begin{equation}\label{PAuser2eq1_2}
u_{ij,T+1}^*=\left\{ \begin{array}{ll}
\!\!1, &\text{if}\; i =\arg\min_{k\in\mathcal D} \beta_{kj}d_{kj}^{m+3}  \\
\!\!0, &\text{otherwise}.
\end{array} \right.
\end{equation}

To obtain the optimal $P_{i,T+1}^*$ from \eqref{dueq1_2}, we derive the first derivative with respect $P_{i,T+1}$ as
\begin{equation}\label{PAuser2eq2_6}
\frac{\partial\mathcal L (P_{i,T+1}, \boldsymbol u_{T+1},  \boldsymbol \beta)}
{\partial P_{i,T+1}}= 1 -\sum_{j\in\mathcal U}\beta_{ij}.
\end{equation}
Note that the optimal $P_{i,T+1}^*=+\infty$ if $1 -\sum_{j\in\mathcal U}d_{ij}<0$ and dual value is $-\infty$. 
To avoid this, we must have $\sum_{j\in\mathcal U}\beta_{ij}\leq 1$.
As a result, we can obtain the optimal solution $P_{i,T+1}^*$ to problem (\ref{optimal_problem5}) as \eqref{PAuser2eq2}.
This completes the proof.

\vspace{-0.3cm}
\subsection{Proof of Lemma 1}
The complexity of the CNN-based illumination distribution encoder and decoder depends on the calculations in convolutional (deconvolutional) layers, max-pooling (unpooling) layers, and a flatten layer.

For each convolutional layer, the calculations based on (\ref{convolution}) is given as:
\vspace{-0.1cm}
\begin{equation}\label{lemma1}
h_{i,j}^{l,m} = f(\sum\limits_{k =1}^ {K_c^{l - 1}} {h_{i,j}^{l - 1,k}w_{1,1}^{l,m} +  \cdots  + h_{i,j + S}^{l - 1,k}w_{1,S}^{l,m} + }  \cdots  + h_{i + S,j + S}^{l - 1,k}w_{S,S}^{l,m} + b_k^{l,m}),
\end{equation}
where $h_{i,j}^{l,m}$ is the element of row $i$ and column $j$ in ${\boldsymbol{H}}_t^{l,m}$, $h_{i,j}^{l - 1,k}$ is the element of row $i$ and column $j$ in ${\boldsymbol{H}}_t^{ l - 1,k}$, $w_{1,1}^{l,m}$ is the element of row $1$ and column $1$ in ${\boldsymbol{W}}_t^{l,m}$, and $b_k^{l,m}$ is the element $k$ of ${\bm{b}}_t^{l,m}$. 
For each $h_{i,j}^{l,m}$, the complexity of calculation is $\mathcal O({K_c^{l-1}}{S^2})$.
Note that, each convolutional layer $l$ consists of $K_c^l$ feature maps and each feature map ${\boldsymbol{H}}_t^{l,m} \in {\mathbb{R}^{\lambda_l \times \lambda_l}}$. Then, we have $i=1, \cdots, \lambda_l$, $j= 1,\cdots, \lambda_l$ and $m=1,\cdots, K_c^l$.
Therefore, the complexity of convolutional layer $l$ is $\mathcal O({\lambda_l}^2{K_c^{l}}{K_c^{l-1}}{S^2})$. 

For each max-pooling layer $l$, the max-pooling operation divides the input feature map ${\boldsymbol{H}}_t^{l-1,m}$ into ${\frac{\lambda_{l-1}^2 }{S_m^2}}$ square areas.
In each $S_m \times S_m$ square area, the max-pooling operation records the most robust feature, whose complexity is $\mathcal O(S_m^2)$.
Hence, the complexity of max-pooling layer $l$ is $\mathcal O({\frac{\lambda_{l-1}^2 }{S_m^2}}S_m^2)=\mathcal O(\lambda_{l-1}^2)$.

For the flatten layer, the flatten operation rewrites input ${\boldsymbol{H}}_t^{L,m}$ to $\bm{x}_t \in {\mathbb{R}^{N}}$, where ${\boldsymbol{H}}_t^{L,m} \in {\mathbb{R}^{\lambda_L \times \lambda_L}}$, $m=1, \cdots, K_c^L$, and $N=\lambda_L^2{K_c^L}$.
Therefore, the complexity of the flatten layer is $\mathcal O(\lambda_L^2{K_c^L})$.

As a result, the complexity of the CNN-based illumination distribution encoder is:
\vspace{-0.1cm}
\begin{equation}
{\mathcal O}\left(\sum\limits_{l = 1}^L {{\lambda _l}^2{K_c^{l}}{K_c^{l-1}}{S^2}}  + \sum\limits_{l = 1}^L {{\lambda _{l - 1}^2} + {\lambda_L^2{K_c^L}}} \right) = {\mathcal O}\left(\sum\limits_{l = 1}^L {{\lambda _l}^2{K_c^{l }}{K_c^{l-1}}{S^2}} \right).
\end{equation}

Due to the symmetry between the CNN-based illumination distribution encoder and the DeCNN-based decoder, the complexity of the decoder is also ${\mathcal O}(\sum\limits_{l = 1}^L {{\lambda _l}^2{K_d^{l}}{K_d^{l-1}}{S^2}} )$.
This completes the proof.

\vspace{-0.3cm}
\subsection{Proof of Lemma 2}
Given representation ${\bm{x}}_t$ for illumination distribution at time slot $t$, the GRU-based predictor extract the temporal characteristics based on (\ref{reset_gate})-(\ref{update_gate}).
For each input ${\bm{x}}_t$, the complexity of reset gate operation in (\ref{reset_gate}) is $\mathcal O\left(N{D_h}+{D_h^2}\right)$, which depends on the size of ${\boldsymbol{W}_r} \in {\mathbb{R}^{{N} \times {D_h}}}$ and ${\boldsymbol{U}_r} \in {\mathbb{R}^{{D_h} \times {D_h}}}$.
Similarly, the complexity of calculating candidate hidden state $\tilde{h}_t^j$ in (\ref{reset}) and the complexity of calculating update gate $z_t^j$ in (\ref{update_gate}) are both $\mathcal O\left(N{D_h}+{D_h^2}\right)$.
The proposed GRU model iteratively updates the hidden states based on (\ref{reset_gate})-(\ref{update_gate}).
Therefore, the complexity of extracting temporal feature for all the input illumination distributions $\boldsymbol{X} = \left(\bm{x}_1, \bm{x}_2,\cdots ,\bm{x}_t, \cdots, \bm{x}_T\right)$ is given as $\mathcal O\left(T \times 3(N{D_h}+{D_h^2})\right)$.
Then, the complexity for the GRU model to output the illumination distribution prediction based on (\ref{output}) is ${\mathcal O}(N{D_h})$, which depends on the size of ${\boldsymbol{W}_o} \in {\mathbb{R}^{{N} \times {D_h}}}$.

Finally, the total complexity of the GRU-based predictor is given as:
\begin{equation}
\setlength{\abovedisplayskip}{2pt}
\setlength{\belowdisplayskip}{2pt}
{\mathcal O}\left( {T(3(N{D_h} + D_h^2)) + N{D_h}} \right)={\mathcal O}\left( {T{D_h}(N + {D_h})} \right).
\end{equation}
This completes the proof.





%



\bibliographystyle{IEEEbib}
\def\baselinestretch{1.35}
\bibliography{illumination}

\end{document}